\documentclass[preprint,aps,floatfix,nofootinbib,a4paper]{revtex4-1}
\pdfoutput=1

\usepackage{amsmath, amssymb, amsfonts, amsthm, latexsym, epsfig, mathrsfs, xcolor, bbm, slashed}

\usepackage[inline]{enumitem}

\usepackage{setspace}
\usepackage[marginal, multiple]{footmisc}

\usepackage[T1]{fontenc}
\usepackage[utf8]{inputenc}
\usepackage{lmodern}

\usepackage[colorlinks, allcolors=blue!70!black, linktocpage]{hyperref}

\numberwithin{equation}{section}

\usepackage{cleveref}

\usepackage{microtype}

\usepackage{floatrow}
\let\OLDtableofcontents\tableofcontents
\renewcommand\tableofcontents[1]{%
    {\baselineskip 0.5ex %
	\OLDtableofcontents{#1}}%
}

\let\OLDthebibliography\thebibliography
\renewcommand\thebibliography[1]{%
	\setstretch{1.079} 
	\OLDthebibliography{#1}%
	\small %
	\setlength{\itemsep}{0.2\baselineskip} 
}

\let\OLDfootnote\footnote
\renewcommand\footnote[1]{%
	\setlength{\footnotesep}{0.75\baselineskip}%
	{\footnotesize \OLDfootnote{#1}}%
}

\setlist[enumerate]{noitemsep, label=(\arabic*), ref=(\arabic*)}

\newlist{condlist}{enumerate}{2}
\setlist[condlist,1]{noitemsep, topsep=0pt, label=(\arabic*), ref=(\arabic*)}
\setlist[condlist,2]{noitemsep, label=(\alph*), ref=(\arabic{condlisti}.\alph*)}
\crefname{condlisti}{condition}{conditions}
\crefname{condlistii}{condition}{conditions}

\newlist{propertylist}{enumerate}{1}
\setlist[propertylist,1]{noitemsep, topsep=0pt, label=(\arabic*), ref=(\arabic*)}
\crefname{propertylisti}{Property}{Properties}

\renewcommand\thesection{\arabic{section}}
\renewcommand\thesubsection{\arabic{subsection}}

\makeatletter
\def\p@subsection{\thesection.}
\def\p@subsubsection{\thesection.\thesubsection.}
\makeatother 

\theoremstyle{plain}

\theoremstyle{definition}
\newtheorem{definition}{Definition}[section]

\theoremstyle{remark}
\newtheorem{remark}{Remark}[section]

\crefname{equation}{Eq.}{Eqs.}
\creflabelformat{equation}{#2#1#3}

\crefname{section}{\S}{\S}
\crefname{appendix}{Appendix}{Appendices}
\crefname{figure}{Fig.}{Figs.}
\crefname{table}{Table}{Tables}

\crefname{definition}{Def.}{Defs.}
\crefname{prop}{Prop.}{Props.}
\crefname{lemma}{Lemma}{Lemmas}
\crefname{corollary}{Cor.}{Cors.}
\crefname{thm}{Theorem}{Theorems}
\crefname{remark}{Remark}{Remarks}

\crefname{ass}{Assumptions}{Assumptions}
\crefname{property}{Properties}{Properties}

\newcommand{\be}{\begin{equation}}
\newcommand{\ee}{\end{equation}}

\newcommand{\lb}{\left}
\newcommand{\rb}{\right}

\newcommand{\mc}{\mathcal}

\newcommand{\ms}{\mathscr}
\newcommand{\mf}{\mathfrak}
\newcommand{\bb}{\mathbb}

\DeclareMathOperator{\sech}{sech}

\newcommand{\set}[1]{\lb\{ #1 \rb\}}
\newcommand{\st}{~:~}

\newcommand{\eqsp}{\, ,\quad} 

\newcommand{\hr}{\begin{center}* * *\end{center}}

\newcommand{\Lie}{\pounds} 
\newcommand{\defn}{\mathrel{\mathop:}=} 

\newcommand{\union}{\cup} 
\newcommand{\inter}{\cap} 

\let\oldsetminus\setminus
\renewcommand{\setminus}{\!\oldsetminus\!} 

\let\oldint\int
\renewcommand{\int}{\oldint\limits}

\let\oldlim\lim
\renewcommand{\lim}{\oldlim\limits}

\renewcommand{\bar}{\overline}

\newcommand{\scri}{\ms I}
\newcommand{\hyp}{\ms H}
\newcommand{\spi}{\text{Spi}}
\newcommand{\cyl}{\ms C}
\newcommand{\nulls}{\ms N}

\newcommand{\dd}[1]{\boldsymbol{#1}} 

\newcommand{\df}[1]{#1} 

\newcommand{\nfrac}[2]{{{}^#1\!\!/\!_#2}}

\newcommand{\half}{\nfrac{1}{2}}

\newcommand{\Omh}{\Omega^\half} 

\newcommand{\rsub}[1]{\scriptscriptstyle{\rm #1}}

\newcommand{\bs}{{\rsub{(BS)}}}

\begin{document}

\setstretch{1.2}


\title{Conservation of asymptotic charges from past to future null infinity: Maxwell fields}
\author{Kartik Prabhu}\email{kartikprabhu@cornell.edu}
\affiliation{Cornell Laboratory for Accelerator-based Sciences and Education (CLASSE),\\ Cornell University, Ithaca, NY 14853, USA}

\begin{abstract}
On any asymptotically-flat spacetime, we show that the asymptotic symmetries and charges of Maxwell fields on past null infinity can be related to those on future null infinity as recently proposed by Strominger. We extend the covariant formalism of Ashtekar and Hansen by constructing a \(3\)-manifold of both null and spatial directions of approach to spatial infinity.  This allows us to systematically impose appropriate regularity conditions on the Maxwell fields near spatial infinity along null directions. The Maxwell equations on this \(3\)-manifold and the regularity conditions imply that the relevant field quantities on past null infinity are antipodally matched to those on future null infinity. Imposing the condition that in a scattering process the total flux of charges through spatial infinity vanishes, we isolate the subalgebra of totally fluxless symmetries near spatial infinity. This subalgebra provides a natural isomorphism between the asymptotic symmetry algebras on past and future null infinity, such that the corresponding charges are equal near spatial infinity. This proves that the flux of charges is conserved from past to future null infinity in a classical scattering process of Maxwell fields. We also comment on possible extensions of our method to scattering in general relativity.
\end{abstract}

\maketitle
\tableofcontents

\section{Introduction}
\label{sec:intro}

In general relativity the asymptotic properties of isolated systems can be studied in three (a priori) \emph{distinct} regimes:
\begin{enumerate*}
    \item at past or future \emph{null infinity} i.e. large separation along null directions either in the past or the future, or,
    \item at \emph{spatial infinity} i.e. large separation along spatial directions.
\end{enumerate*}
It is well-known that at (both past or future) null infinity one obtains an infinite-dimensional asymptotic symmetry group --- the Bondi-Metzner-Sachs (BMS) group --- along with the corresponding ``conserved'' charges and fluxes due to gravitational radiation \cite{BBM, Sachs1, Sachs2, Penrose, GW, AS-symp, WZ}. Similarly, the asymptotic structure at spatial infinity has been analysed, in a \((3+1)\)-formalism \cite{ADM, ADMG, CR} and in a \(4\)-dimensional formalism \cite{Beig-Schmidt, AH, Sommers, Ash-in-Held, Ash-Rom, Friedrich, AES}. In this case, one again obtains an infinite-dimensional asymptotic symmetry group --- the Spi-group --- and conserved charges corresponding to the Arnowitt-Deser-Misner (ADM) energy and angular momentum. For a detailed review of asymptotic structures in general relativity see \cite{Geroch-asymp}.

Very little is known about the relation between the symmetries and charges defined independently in these three regimes (see \cite{Ash-Mag-Ash, AS-ang-mom, HL-GR-matching} for some known results). For example, in the context of a gravitational scattering problem one could ask: Does the flux of charges for some symmetry on past null infinity \(\scri^-\) equal the flux of charges for a ``corresponding'' symmetry on future null infinity \(\scri^+\)? Any attempt to answer this question would first need some appropriate notion of ``corresponding'' i.e. some isomorphism between the asymptotic symmetries on past null infinity to the ones on future null infinity. Then, using the evolution equations for the fields one would have to show that the incoming fields prescribed on past null infinity evolve to future null infinity so that the fluxes of the charges are equal.

This ``matching'' problem has received renewed interest due to the recent conjecture by Strominger \cite{Stro-CK-match} that all the asymptotic symmetries given by the BMS group on \(\scri^-\) can be related to the ones on \(\scri^+\) through an antipodal reflection on suitable limiting cross-sections near spatial infinity. Such a matching gives a \emph{global} ``diagonal'' asymptotic symmetry group for general relativity. If such a diagonal symmetry group can be found, along with similar matching conditions for the gravitational field quantities on \(\scri^-\) to those on \(\scri^+\),\footnote{Such matching conditions on the gravitational fields were assumed as ``boundary conditions'' in \cite{Stro-CK-match}.} it would imply that the fluxes of the corresponding charges will be \emph{globally} conserved in the sense that the incoming fluxes at past null infinity would equal the outgoing fluxes at future null infinity.\footnote{This inherently assumes that appropriate conditions are satisfied near timelike infinities and on event horizons, if any exist; see \cref{rem:timelike-infinity}.} It has been further conjectured that this diagonal group is a symmetry of the scattering matrix in quantum gravity \cite{Stro-CK-match} and the corresponding flux conservation laws have been related to various soft theorems \cite{HLMS, SZ}, and also speculated to play a role in the black hole information loss problem \cite{HPS, Hawking} (see \cite{BP} for a contrarian view).

However, the validity of such matching conditions for the asymptotic symmetries and charges has not been proven even in classical general relativity. The argument for the matching conditions given in \cite{Stro-CK-match} for the Christodoulou-Klainerman class of spacetimes (CK-spacetimes) \cite{CK} is not sufficient since, as shown by Ashtekar \cite{Ash-CK-triv}, the charges for all the angle-dependent supertranslations vanish in the limit to spatial infinity in such spacetimes. The main obstacle to resolving the matching problem is as follows. Given a physical spacetime we can define a conformally-related unphysical spacetime, the Penrose conformal-completion, to study the asymptotic behaviour. In the unphysical spacetime, null infinities \(\scri^\pm\) are smooth null boundaries while spatial infinity is a boundary point \(i^0\) acting as the vertex of ``the light cone at infinity''. On Minkowski spacetime the unphysical spacetime is smooth (in fact, analytic) at \(i^0\). So we can easily identify the null generators (and fields) on \(\scri^-\) with those on \(\scri^+\) by ``passing through'' \(i^0\). However, in more general spacetimes, the unphysical metric is not even once-differentiable at spatial infinity unless the ADM mass of the spacetime vanishes \cite{AH}, and the unphysical spacetime manifold \emph{does not} have a smooth differential structure at \(i^0\). Thus the identification between the null generators of \(\scri^-\) and \(\scri^+\), and the corresponding fields, becomes much more difficult.

As a simpler problem such matching conditions can be investigated for linearised gravity or Maxwell fields on a fixed background asymptotically-flat spacetime (we comment on the case of full general relativity in \cref{sec:disc}). Even in this case the problem has only been resolved on Minkowski spacetime. The antipodal matching of all the infinite number of symmetries and charges has been shown, for Maxwell fields by Campiglia and Eyheralde \cite{CE} (see also \cite{HT-em}; and the generalisation to \(p\)-form gauge fields \cite{AES-p-form}), and for supertranslations in linearised gravity by Troessaert \cite{Tro}.\footnote{The result of \cite{Tro} can be viewed as the linearisation around a Minkowski background of the more general analysis in \cite{HL-GR-matching}.} The key improvement in these works over \cite{Stro-CK-match} is that the antipodal matching of the relevant fields is not assumed a priori as a boundary condition, but follows from the equations of motion and regularity conditions near spatial infinity. However, these proofs rely on an asymptotic expansion of the fields in suitable coordinates near both null and spatial infinity on Minkowski spacetime. Explicitly transforming the various fields from one set of coordinates to another then gives the sought after matching conditions. But in more general spacetimes such ``nice'' coordinates and their explicit transformations are not available near \(i^0\). Thus, in general asymptotically-flat spacetimes the matching problem, even for Maxwell fields, has not been resolved.

The main goal of this paper is to prove the matching conditions for \emph{all} asymptotic symmetries of Maxwell fields on \emph{any} asymptotically-flat spacetime.\footnote{For the usual Coulomb charge one can use the method of \cite{Ash-Mag-Ash}.} We will use the definition of asymptotic-flatness for the background spacetime and the Maxwell fields given by Ashtekar and Hansen \cite{AH, Ash-in-Held} to treat both null and spatial infinities in a unified spacetime-covariant manner (\cref{def:AH}). In the Ashtekar-Hansen formalism, instead of working directly at the point \(i^0\) where sufficiently smooth structures are unavailable, one works on the space of spatial directions along which we can approach \(i^0\). This space of spatial directions at \(i^0\) is a timelike-unit-hyperboloid \(\hyp\) in the tangent space at \(i^0\) (\cref{fig:hyperboloid}). Suitably conformally rescaled Maxwell fields, whose limits to \(i^0\) depend on the direction of approach, induce \emph{smooth} fields on \(\hyp\) and we can study these smooth limiting fields using standard tools. The asymptotic symmetries at spatial infinity then give us infinitely many charges on \(\hyp\) in terms of these smooth limiting fields.

The Ashtekar-Hansen formalism only specifies the behaviour of fields as they approach \(i^0\) along spatial directions and no conditions are imposed approaching \(i^0\) along the null infinity. For the matching problem we are interested in precisely the behaviour of fields at \(i^0\) along \(\scri^\pm\). Thus, we augment the Ashtekar-Hansen formalism by constructing a space of both null and spatial directions of approach to \(i^0\). This space is a cylinder \(\cyl\) in the tangent space at \(i^0\), which is diffeomorphic to a conformal-completion of \(\hyp\),\footnote{Note, this additional conformal-completion of \(\hyp\) does not arise directly from the standard conformal-completion of the physical spacetime (details in \cref{sec:null-dir}).} with two boundaries \(\nulls^\pm\) corresponding to the directions of approach to \(i^0\) in null directions along \(\scri^\pm\) (\cref{fig:cylinder}). Using this diffeomorphism, we can study the Maxwell fields and symmetries on \(\cyl\), instead of on \(\hyp\), and consider the limits of Maxwell fields as they approach \(i^0\) in both null and spatial directions.

We can then ask about two different limits of the Maxwell fields and symmetries:
\begin{enumerate*}
    \item first take the limit to \(\scri^\pm\) and then towards \(i^0\), or,
    \item first take the limit to \(i^0\) along spatial directions (now represented by \(\cyl\setminus\nulls^\pm\)) and then take the limit where the direction of approach becomes null.
\end{enumerate*}
In general, neither of these limits might exist given the conditions by Ashtekar and Hansen. Thus, we impose additional \emph{null-regularity} conditions on (some components of) the Maxwell fields (\cref{def:F-null-regular}). These conditions act as ``continuity'' conditions on the Maxwell fields at \(i^0\), implying that both limits, taken as described above, exist and the induced limiting fields on the boundaries \(\nulls^\pm\) obtained by both limiting procedures match. The null-regularity conditions further ensure the physical requirement that the radiated flux of charges through \(\scri^\pm\) is finite in any scattering process. This will lead us to a partial matching of the symmetries, whereby any symmetry on \(\scri^\pm\) gives some (not unique) symmetry on \(\cyl\) such that they match continuously at \(\nulls^\pm\).

Using the Maxwell equations on \(\cyl\) we show that, with our null-regularity conditions, the fields entering the expression for the charges from the past null directions \(\nulls^-\) match antipodally to those from the future null directions \(\nulls^+\). Finally, we can isolate a subalgebra of symmetries such that the total flux of charges across \emph{all} of \(\cyl\) corresponding to these symmetries vanishes. This corresponds to the physical requirement that in a scattering process one only is concerned with the fields on null infinity, and any flux through spatial infinity is ``non-dynamical''. We emphasise that this is not a restriction on the kinds of fields we consider (unlike the previously discussed null-regularity conditions), but a choice of symmetries relevant to a scattering process. Such totally fluxless symmetries on \(\cyl\) then give us the desired isomorphism between the symmetries and a conservation law for the fluxes between \(\scri^-\) and \(\scri^+\), proving the conjecture in \cite{Stro-CK-match} for Maxwell fields on any asymptotically-flat spacetime.

\hr

The rest of the paper is organised as follows. In \cref{sec:AH} we review the Ashtekar-Hansen structure of spactimes that are asymptotically-flat at both null and spatial infinity, and the asymptotic symmetries and charges for electromagnetic fields on such spacetimes. In \cref{sec:null-dir}, we summarise the construction of the space \(\cyl\) at spatial infinity that includes both null and spatial directions and its relation to the unit-hyperboloid \(\hyp\) in the Ashtekar-Hansen framework. In \cref{sec:Maxwell-partial-matching}, we impose suitable regularity conditions on the Maxwell fields on null infinity which ensure that charges defined on null infinity remain finite as we approach spatial infinity; this allows us to match the symmetries on null infinity to the ones at spatial infinity. We introduce the reduced algebra of symmetries such that the corresponding total flux of charges across spatial infinity vanishes; this then gives us the antipodal matching conditions, the global diagonal symmetry algebra and the flux conservation between past and future null infinity. We end with \cref{sec:disc} summarising our results and discussing possible extensions to general relativity. We collect the definitions of direction-dependent differential structures and tensors in \cref{sec:dd}. In \cref{sec:Sigma}, we show that the rescaling function used to construct the space \(\cyl\) of null and spatial directions at spatial infinity exists. In \cref{sec:BS}, we relate our covariant approach to the coordinate-based approaches used on Minkowski spacetime. We analyse the solutions of Maxwell equations at spatial infinity in \cref{sec:wave-hyp}.


\hr

We use an abstract index notation with indices \(a,b,c,\ldots\) for tensor fields. Quantities defined on the physical spacetime will be denoted by a ``hat'', while the ones on the conformally-completed unphysical spacetime are without the ``hat'' e.g. \(\hat g_{ab}\) is the physical metric while \(g_{ab}\) is the unphysical metric on the conformal-completion. We will raise and lower indices on tensors with \(g_{ab}\) and explicitly write out \(\hat g_{ab}\) when used to do so. We denote directions at \(i^0\) by an overhead arrow e.g. \(\vec N\) denotes directions which are either null or spatial while \(\vec\eta\) denotes spatial directions. Regular direction-dependent limits of tensor fields will be denoted by a boldface symbol e.g. \(\dd F_{ab}(\vec\eta)\) is the limit of the (rescaled) Maxwell field along spatial directions at \(i^0\). We collect our conventions on the orientations of the normals defined on various manifolds in \cref{tab:orientation}.

\begin{table}[!h]
\centering
\begin{tabular}{|l|l|}
	\hline
	 Normal vector field & Orientation \\ \hline
     \(n^a\) & null and future-pointing at \(\scri^+\), past-pointing at \(\scri^-\) \\
     \(l^a\) & null and future-pointing at \(\scri^+\), past-pointing at \(\scri^-\) \\
     \(\dd\eta^a\) & spatial and inward-pointing at \(i^0\) \\
     \(\dd u^a\) & timelike and future-pointing at some cross-section \(S\) of \(\hyp\) \\
     \(\dd\Sigma^{-1}\dd U^a\) & timelike and future-pointing at \(\nulls^+\), past-pointing at \(\nulls^-\) on \(\cyl\) \\
	\hline
\end{tabular}
\caption{Conventions for orientation of normals}
\label{tab:orientation}
\end{table}

\section{Asymptotic-flatness at null and spatial infinity: Ashtekar-Hansen structure}
\label{sec:AH}

We define spacetimes which are asymptotically-flat at null and spatial infinity following \cite{AH, Ash-in-Held}. To distinguish this from other constructions, we refer to this as an Ashtekar-Hansen structure. We use the following the notation for causal structures from \cite{Hawking-Ellis}: \(J(i^0)\) is the causal future of a point \(i^0\) in \(M\), \(\bar J(i^0)\) is its closure, \(\dot J(i^0) \) is its boundary and \(\scri \defn \dot J(i^0) - i^0\). We also use the definition and notation for direction-dependent tensors from \cref{sec:dd}.

\begin{definition}[Ashtekar-Hansen structure \cite{Ash-in-Held}]\label{def:AH}
	A \emph{physical} spacetime \((\hat M, \hat g_{ab})\) has an \emph{Ashtekar-Hansen structure} if there exists another \emph{unphysical} spacetime \((M, g_{ab})\), such that
\begin{condlist}
	\item \(M\) is \(C^\infty\) everywhere except at a point \(i^0\) where it is \(C^{>1}\),
	\item the metric \(g_{ab}\) is \(C^\infty\) on \(M-i^0\), and \(C^0\) at \(i^0\) and \(C^{>0}\) along spatial directions at \(i^0\)
	\item there is an embedding of \(\hat M\) into \(M\) such that \(\bar J(i^0) = M - \hat M\),
	\item there exists a function \(\Omega\) on \(M\), which is \(C^\infty\) on \(M-i^0\) and \(C^2\) at \(i^0\) so that \(g_{ab} = \Omega^2 \hat g_{ab}\) on \(\hat M\) and
		\begin{condlist}
			\item \(\Omega = 0\) on \(\dot J(i^0)\) 
			\item \(\nabla_a \Omega \neq 0\) on \(\scri\)
			\item at \(i^0\), \(\nabla_a \Omega = 0\), \(\nabla_a \nabla_b \Omega = 2 g_{ab}\) \label{cond:Omega-at-i0}
		\end{condlist}
	\item There exists a neighbourhood \(N\) of \(\dot J(i^0)\) such that \((N, g_{ab})\) is  strongly causal and time orientable, and in \(N \inter \hat M\) the physical metric \(\hat g_{ab}\) satisfies the vacuum Einstein equation \(\hat R_{ab} = 0\),
	\item The space of integral curves of \(n^a = g^{ab}\nabla_b \Omega\) on \(\dot J(i^0)\) is diffeomorphic to the space of null directions at \(i^0\). \label{cond:int-curves}
    \item The vector field \(\varpi^{-1} n^a\) is complete on \(\scri\) for any smooth function \(\varpi\) on \(M - i^0\) such that \(\varpi > 0\) on \(\hat M \union \scri\) and \(\nabla_a(\varpi^4 n^a) = 0\) on \(\scri\). \label{cond:complete}
    \end{condlist}
\end{definition}
One can also use much weaker differentiability requirements on \(M-i^0\) and also consider non-vacuum spacetimes (see \cite{AH, Ash-in-Held}), but we choose not to do so for simplicity.

The physical role of the conditions in \cref{def:AH} are explained in \cite{Ash-in-Held}. In particular, these conditions  imply that
\begin{enumerate*}[label=(\roman*)]
	\item The point \(i^0\) is spacelike related to all points in the physical spacetime \(\hat M\), and represents \emph{spatial infinity}.
	\item \(\scri \defn \dot J(i^0) - i^0\) consists of two disconnected pieces --- the future piece \(\scri^+\) and the past piece \(\scri^-\) --- which are both smooth null submanifolds of \(M\), representing future and past \emph{null infinities}, respectively.
\end{enumerate*}
Note that the metric \(g_{ab}\) is only \(C^{>0}\) at \(i^0\) along spatial directions, that is, the metric is continuous but the metric connection (or Christoffel symbols in some \(C^{>1}\) coordinate chart, see \cref{sec:dd}) is allowed to have limits which depend on the direction of approach to \(i^0\). As mentioned in the Introduction this low differentiability structure is necessary to include spacetimes with non-vanishing ADM mass, and can be roughly visualised as the gravitational ``lines of force'' all converge to a single point at \(i^0\) and thus the metric connection cannot be smooth there. Herberthson \cite{Herb-Kerr} has shown that the Kerr family of spacetimes satisfy \cref{def:AH}, in fact, the unphysical metric for Kerr spacetimes is \(C^{>0}\) in \emph{both} null and spatial directions at \(i^0\). In the following we only require that the unphysical metric is \(C^0\) along null directions at \(i^0\).

For a given physical spacetime \((\hat M, \hat g_{ab})\), the choice of an Ashtekar-Hansen structure is not unique. There is an ambiguity in the choice of the \(C^{>1}\) differential structure at \(i^0\) given by a \(4\)-parameter family of \emph{logarithmic translations} which simultaneously change the \(C^{>1}\)-structure and the conformal factor at \(i^0\) \cite{Berg, Ash-log, Chr-log}. Given a choice of the \(C^{>1}\)-structure the additional ambiguity in the choice of the conformal factor \(\Omega\) is as follows.
\begin{remark}[Freedom in the conformal factor \cite{AH,Ash-in-Held}]
\label{rem:freedom-Omega}
The freedom in the choice of the conformal factor in \cref{def:AH} is given by \(\Omega \mapsto \omega\Omega\) where the function \(\omega\) satisfies
\begin{propertylist}
    \item \(\omega > 0\) on \(M\) 
    \item \(\omega\) is smooth on \(M - i^0\)
    \item \(\omega\) is \(C^{>0}\) at \(i^0\) and \(\omega \vert_{i^0} = 1\)
\end{propertylist}
\end{remark}

In the following we will work with a fixed the unphysical spacetime given by some choice of the \(C^{>1}\)-structure at \(i^0\) and some choice of conformal factor \(\Omega\). In the end, we argue that our results are independent of these choices (see \cref{rem:change-Sigma}). We however note that, all spacetimes satisfying \cref{def:AH} have the same metric at \(i^0\), that is, the unphysical metric \(g_{ab}\) at \(i^0\) is \emph{universal} and cannot even be further conformally-rescaled (since \(\omega\vert_{i^0} = 1\)) \cite{AH}.

\hr

The Ashtekar-Hansen structure \cref{def:AH} gives us the standard structure on null infinity \(\scri\). It follows from the Einstein equation that the vector field \(n^a\) is a null geodesic generator of \(\scri^\pm \cong \bb R \times \bb S^2\), which is future/past pointing on \(\scri^\pm\) respectively. Note that \(n^a\) is \emph{need not} be affinely-parameterised, i.e. \(n^b\nabla_b n^a \vert_{\scri^\pm} \neq 0\) in the choice of conformal factor allowed by \cref{cond:Omega-at-i0}. The function \(\varpi\) in \cref{cond:complete} used to define a complete divergence-free normal \(\varpi^{-1}n^a\) cannot be used as a conformal-rescaling at \(i^0\) since \(\varpi\) will diverge at \(i^0\) (see footnote 2, \S~11.1 \cite{Wald-book} and \cref{sec:BS}). In particular, the unphysical metric in the Bondi conformal frame and the corresponding Bondi-Sachs coordinates are ill-behaved near \(i^0\).

 Denote by \(q_{ab}\) the pullback of \(g_{ab}\) to \(\scri\). This defines a degenerate metric on \(\scri\) with \(q_{ab} n^b = 0\). It is convenient to introduce some foliation of \(\scri\) by a one-parameter family of cross-sections diffeomorphic to \(\bb S^2\). The pullback of \(q_{ab}\) to any cross-section \(S\) defines a Riemannian metric on \(S\). Then, for any choice of foliation, there is a unique \emph{auxilliary normal} vector field \(l^a\) at \(\scri\) such that
\be\label{eq:l-props}
    l^a l_a = 0 \eqsp l^a n_a = -1 \eqsp q_{ab}l^b = 0
\ee
In our conventions, \(l^a\) is future/past pointing at \(\scri^\pm\), respectively (\cref{tab:orientation}). We further have
\be\label{eq:null-fields}
    q_{ab} = g_{ab} + 2 n_{(a} l_{b)} \eqsp \varepsilon_{abc} = l^d \varepsilon_{dabc} \eqsp \varepsilon_{ab} = n^c \varepsilon_{cab}
\ee
where \(\varepsilon_{abc}\) defines a volume element on \(\scri\) and \(\varepsilon_{ab}\) is the area element on any cross-section \(S\) of the foliation corresponding to the metric \(q_{ab}\).\\

Since spatial infinity is represented by a single point \(i^0\) in \(M\), various physical fields of interest, in general, will not be continuous at \(i^0\) but will admit \emph{direction-dependent} limits. Hence it is rather inconvenient to study such fields directly at the point \(i^0\). This can be remedied using the technique of a \emph{blowup} from algebraic geometry (see \cite{Harris}). Instead of working at the point \(i^0\) one works on the space of directions at \(i^0\) (the blowup).\footnote{To be precise, the blowup consists of a bundle over the space of directions (see \cref{rem:spi} for the case at spatial infinity), but we will use this terminology more loosely for the space of directions itself.} The fields which have regular direction-dependent limits to \(i^0\) (as defined in \cref{sec:dd}) induce smooth tensor fields on the space of directions. The space of spatial directions at \(i^0\) was constructed in \cite{AH}, we review the aspects of this construction relevant for our analysis. Later in \cref{sec:null-dir}, we will construct a closely related blowup which includes the null directions at \(i^0\) and will be more suitable for relating fields at spatial infinity to those on null infinity.

\begin{figure}[h!]
	\centering
	\includegraphics[width=0.3\textwidth]{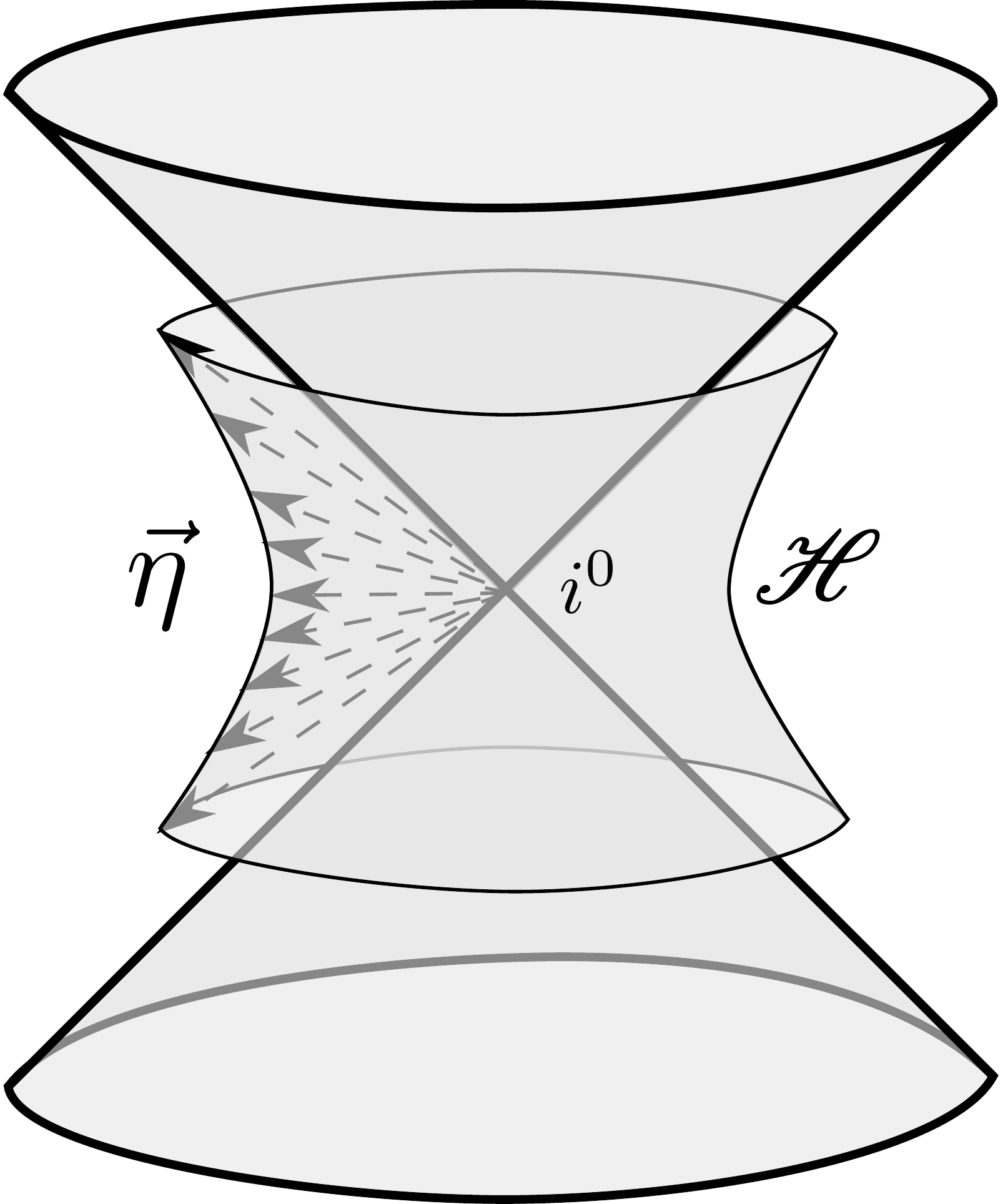}
	\caption{The unit-hyperboloid \(\hyp\) in \(Ti^0\) representing spatial directions \(\vec\eta\) at \(i^0\) (drawn as dashed rays). Note that \(\hyp\) is a \emph{non-compact} manifold without boundary and we have only drawn a part for illustration.}
    \label{fig:hyperboloid}
\end{figure}

Along spatial directions \(\nabla_a \Omh\) is \(C^{>-1}\) at \(i^0\) and 
\be\label{eq:eta-defn}
    \dd\eta^a \defn \lim_{\to i^0} \nabla^a \Omh
\ee
determines a \(C^{>-1}\) spatial unit vector field at \(i^0\) representing the spatial directions \(\vec\eta\) at \(i^0\). The space of such spatial directions in \(Ti^0\) is a unit-hyperboloid \(\hyp\) (see \cref{fig:hyperboloid}) 

If \(T^{a \ldots}{}_{b \ldots}\) is a \(C^{>-1}\) tensor field at \(i^0\) in spatial directions then, \(\lim_{\to i^0} T^{a \ldots}{}_{b \ldots} = \dd T^{a \ldots}{}_{b \ldots}(\vec\eta)\) is a smooth tensor field on \(\hyp\). Further, the derivatives of \(\dd T^{a \ldots}{}_{b \ldots}(\vec\eta)\) to all orders with respect to the direction \(\vec\eta\) satisfy\footnote{The factors of \(\Omh\) on the right-hand-side of \cref{eq:dd-der-spatial} convert between \(\nabla_a\) and the derivatives with respect to the directions \cite{Ash-in-Held,Geroch-asymp}; see also \cref{sec:dd}.}
\be\label{eq:dd-der-spatial}
    \dd \partial_c \cdots \dd \partial_d \dd T^{a \ldots}{}_{b \ldots}(\vec\eta) = \lim_{\to i^0} \Omh \nabla_c \cdots \Omh \nabla_d T^{a \ldots}{}_{b \ldots}
\ee
where \(\dd \partial_a\) is the derivative with respect to the directions \(\vec \eta\) defined by 
\be\label{eq:dd-derivative-spatial}\begin{split}
    \dd v^c \dd \partial_c \dd T^{a \ldots}{}_{b \ldots}(\vec\eta) & \defn \lim_{\epsilon \to 0} \frac{1}{\epsilon} \big[ \dd T^{a \ldots}{}_{b \ldots}(\vec\eta + \epsilon \vec v) - \dd T^{a \ldots}{}_{b \ldots}(\vec\eta) \big] \quad \text{for all } \dd v^a \in T\hyp \\
    \dd \eta^c \dd \partial_c \dd T^{a \ldots}{}_{b \ldots}(\vec\eta) & \defn 0
\end{split}\ee
It can be checked that, along spatial directions, \cref{eq:dd-der-spatial} is equivalent to the definition \cref{eq:dd-derivative} given in terms of a \(C^{>1}\) coordinate chart.

The metric \(\dd h_{ab}\) induced on \(\hyp\) by the universal metric \(\dd g_{ab}\) at \(i^0\), satisfies
\be\label{eq:d-eta-h}
    \dd h_{ab} \defn \dd g_{ab} - \dd \eta_a \dd \eta_b = \dd \partial_a \dd \eta_b
\ee
Further, if \(\dd T^{a \ldots}{}_{b \ldots}(\vec\eta)\) is orthogonal to \(\dd\eta^a\) in all its indices then it defines a tensor field \(\dd T^{a \ldots}{}_{b \ldots}\) intrinsic to \(\hyp\). Then we can project \emph{all} the indices in \cref{eq:dd-der-spatial} using \(\dd h_{ab}\) to define a derivative operator \(\dd D_a\) intrinsic to \(\hyp\) which is also the covariant derivative operator associated to \(\dd h_{ab}\).\footnote{This follows from \cref{eq:d-eta-h}, and \(\dd\partial_c \dd g_{ab} = 0\) since \(\dd g_{ab}\) is continuous and hence direction-independent at \(i^0\).} We also define
\be\label{eq:volume-hyp}
    \dd\varepsilon_{abc} \defn - \dd\eta^d \dd\varepsilon_{dabc} \eqsp \dd\varepsilon_{ab} \defn \dd u^c \dd\varepsilon_{cab}
\ee
where \(\dd\varepsilon_{abcd}\) is volume element at \(i^0\) correspoding to the metric \(\dd g_{ab}\), \(\dd\varepsilon_{abc}\) is the induced volume element on \(\hyp\), and \(\dd\varepsilon_{ab}\) is the induced area element on some cross-section \(S\) of \(\hyp\) with a future-pointing timelike normal \(\dd u^a\) such that \(\dd h_{ab} \dd u^a \dd u^b = -1\). 

We note that \(\hyp\) admits a reflection isometry as follows. On the unit-hyperboloid we can introduce coordinates \((\tau, \theta^A)\) --- where \(\tau \in (-\infty,\infty)\), and \(\theta^A = (\theta,\phi )\) are the standard coordinates on \(\bb S^2\) with \(\theta \in [0,\pi]\) and  \(\phi \in [0 ,2\pi)\) --- so that the metric on \(\hyp\) is
\be
	\dd h_{ab} \equiv - d\tau^2 + \cosh^2\tau (d\theta^2 + \sin^2\theta d\phi^2 )
\ee
Note that this is possible only because \(\dd g_{ab}\) is universal at \(i^0\) and hence \(\dd h_{ab}\) is always a unit-hyperboloid metric. The metric \(\dd h_{ab}\) has a reflection isometry \(\Upsilon\) 
\be\label{eq:reflection-hyp}\begin{split}
    \Upsilon &: \hyp \to \hyp : (\tau, \theta^A) \mapsto (-\tau, -\theta^A) \\
    \text{with } \Upsilon &\circ \dd h_{ab} = \dd h_{ab}
\end{split}\ee
which is obtained by composing the time-reflection \(\tau \mapsto - \tau\) with the antipodal reflection \(\theta^A = (\theta, \phi) \mapsto - \theta^A = (\pi - \theta, \phi\pm \pi)\) on \(\bb S^2\) (where the sign is chosen so that \(\phi \pm \pi \in [0,2\pi)\)). We have denoted the natural action of the reflection map \(\Upsilon\) on tensor fields on \(\hyp\) by \(\Upsilon \circ\). \\

\begin{remark}[The \(\spi\) manifold]
\label{rem:spi}
The Ashtekar-Hansen structure also provides us with an additional universal structure at \(i^0\) given by a \(\bb R\)-principal bundle over \(\hyp\), called \(\spi\) \cite{AH}. This is useful in the definition of asymptotic symmetries and charges (in particular the \(\spi\)-supertranslations) at spatial infinity in general relativity. For the Maxwell case, which we consider in this paper, we will not need this additional structure.
\end{remark}

\subsection{Asymptotic symmetries and charges for Maxwell fields}
\label{sec:Maxwell}

In the physical spacetime \(\hat M\), let \(\hat F_{ab}\) be the Maxwell field tensor and \(\hat J^a\) the Maxwell charge-current, satisfying the Maxwell equations 
\be\label{eq:Max-phys}
    \hat g^{ac} \hat g^{bd}\hat\nabla_b \hat F_{dc} = \hat J^a \eqsp \hat\nabla_{[a} \hat F_{bc]} = 0
\ee
In the unphysical spacetime \(M\) with \(F_{ab} \defn \hat F_{ab}\) and \(J^a \defn \Omega^{-4}\hat J^a\) we have 
\be\label{eq:Max}
    \nabla_b F^{ba} = J^a \eqsp \nabla_{[a} F_{bc]} = 0
\ee

At null infinity, we assume the standard asymptotic conditions that
\be\label{eq:F-asymp-null}
    F_{ab} \text{ and } J^a \text{ have smooth limits to } \scri    
\ee
The asymptotic symmetries at null infinity are given by functions \(\lambda\) which have smooth limits to \(\ms \scri\) such that \(\Lie_n \lambda = 0\) on \(\scri\). Thus, the asymptotic symmetry algebra at \(\scri^\pm\), respectively, is
\be\label{eq:symm-null}
    \mf g^\pm = \set{ \lambda \in C^\infty(\scri^\pm) \st \Lie_n \lambda = 0  } \cong C^\infty(\bb S^2)
\ee
On some cross-section \(S\) of \(\scri\) each asymptotic symmetry \(\lambda\) has corresponding \emph{electric} and \emph{magnetic} charges given by the pullback of \((*F)_{ab}\) and \(F_{ab}\), respectively, to \(S\). In our conventions these are
\be\label{eq:charge-null}\begin{split}
	\mc Q_{\rsub{(e)}}[\lambda; S] & = - \int_S \df\varepsilon_2~ \lambda F_{ab}l^a n^b \\
    \mc Q_{\rsub{(m)}}[\lambda; S] & = \int_S \df\varepsilon_2~ \lambda (*F)_{ab}l^a n^b = \tfrac{1}{2} \int_S \df\varepsilon_2~ \lambda F_{ab} \varepsilon^{ab} 
\end{split}\ee
where \(\df\varepsilon_2 \equiv \varepsilon_{ab}\) is the area element on \(S\) (\cref{eq:null-fields}). These include the usual Coloumb charges for \(\lambda = \text{constant}\) as well as the charges for ``angle-dependent'' symmetries, which are the analogues of supertranslations in general relativity. The flux of these charges in the region \(\Delta\scri\) of \(\scri\) bounded by two cross-sections \(S_2\) and \(S_1\) is given by the exterior derivative of the integrand in \cref{eq:charge-null}. Using \cref{eq:null-fields} and the Maxwell equations \cref{eq:Max} we get
\be\label{eq:flux-null}\begin{split}
    \mc F_{\rsub{(e)}}[\lambda; \Delta\scri] &\defn \mc Q_{\rsub{(e)}}[\lambda; S_2] - \mc Q_{\rsub{(e)}}[\lambda; S_1] = \int_{\Delta\scri} \df\varepsilon_3~ \lb(\lambda n_a J^a + q^{ab} \nabla_a\lambda \mc E_b \rb) \\
    \mc F_{\rsub{(m)}}[\lambda; \Delta\scri] &\defn \mc Q_{\rsub{(m)}}[\lambda; S_2] - \mc Q_{\rsub{(m)}}[\lambda; S_1] = \int_{\Delta\scri} \df\varepsilon_3~ \varepsilon^{ab} \nabla_a\lambda \mc E_b 
\end{split}\ee
where \(\mc E_a\) is the pullback of \(F_{ab}n^b\) to \(\scri\), and \(\df\varepsilon_3 \equiv \varepsilon_{abc}\) is the volume element on \(\scri\) (\cref{eq:null-fields}). Note, due to our orientation conventions (\cref{tab:orientation}) in \cref{eq:flux-null}, \(S_2\) is in the future of \(S_1\) and the fluxes measure the incoming flux on both \(\scri^{\pm}\).\\

Following \cite{AH,Ash-in-Held}, at \(i^0\) we assume the asymptotic conditions that along spatial directions
\be\label{eq:F-asymp-spatial}
    \lim_{\to i^0} \Omega F_{ab} = \dd F_{ab}(\vec\eta) \text{ is } C^{>-1} \eqsp \lim_{\to i^0} \Omega^{\nfrac{3}{2}} J^a = 0
\ee
Then \(\dd F_{ab}(\vec\eta)\) is completely determined by the \emph{electric} and \emph{magnetic} fields\footnote{Note that the electric and magnetic decomposition is with respect to the \emph{timelike} surface \(\hyp\).}
\be\label{eq:E-B-defn}\begin{split}
    & \dd E_a (\vec\eta) \defn \dd F_{ab} (\vec\eta) \dd\eta^b \eqsp \dd B_a (\vec\eta) \defn (* \dd F)_{ab}(\vec\eta) \dd\eta^b \\[1.5ex]
    \text{with } & \dd F_{ab}(\vec\eta) = -2 \dd\eta_{[a} \dd E_{b]}(\vec\eta) + \dd\eta^d \dd\varepsilon_{dabc} \dd B^c (\vec\eta)
\end{split}\ee
where \((*\dd F)_{ab}(\vec\eta) \defn \tfrac{1}{2} \dd\varepsilon^{cd}{}_{ab}\dd F_{cd}(\vec\eta)\) is the Hodge dual with respect to the unphysical volume element \(\dd\varepsilon_{abcd}\) at \(i^0\). The electric and magnetic fields are orthogonal to \(\dd\eta^a\) and thus induce intrinsic fields \(\dd E_a\) and \(\dd B_a\) on \(\hyp\). Multiplying the Maxwell equations \cref{eq:Max} by \(\Omega^{\nfrac{3}{2}}\) and taking the limit to \(i^0\) in spatial directions we get the asymptotic equations on \(\hyp\) (see \cite{AH} for details)
\begin{subequations}\label{eq:Max-hyp}\begin{align}
    \dd D^a \dd E_a = \dd D^a \dd B_a = 0  \label{eq:Max-hyp-div} \\ 
    \dd D_{[a} \dd E_{b]} = \dd D_{[a} \dd B_{b]} = 0 \label{eq:Max-hyp-curl}
\end{align}\end{subequations}
We defer the analysis of solutions to these equations to \cref{sec:wave-hyp}.

The asymptotic symmetries at spatial infinity are given by functions \(\lambda\) which are \(C^{>-1}\) in spatial directions at \(i^0\) so that \(\lim_{\to i^0}\lambda = \dd\lambda(\vec\eta)\). Thus, at spatial infinity the asymptotic symmetry algebra \(\mf g^0\) is given by all smooth functions \(\dd\lambda\) on \(\hyp\) induced by \(\dd\lambda(\vec\eta)\) i.e.
\be\label{eq:symm-hyp}
    \mf g^0 \cong C^\infty(\hyp)
\ee
Consider a cross-section \(S\) of \(\hyp\), with unit future-pointing timelike normal \(\dd u^a\). The electric and magnetic charges for \(\dd\lambda \in \mf g^0\) on \(S\) are\footnote{Cross-sections of \(\hyp\) represent the limits to \(i^0\) of \(2\)-spheres on a spacelike Cauchy surface of the physical spacetime \(\hat M\) which extends to a spacelike \(C^{>1}\) surface at \(i^0\) in the unphysical spacetime \(M\). The charges in \cref{eq:charge-hyp} are then the direction-dependent limits of the charges evaluated on such \(2\)-spheres. The electric charge in \cref{eq:charge-hyp} has a negative sign since \(\dd\eta^a\) is inward-pointing; see \cref{tab:orientation}.}
\be\label{eq:charge-hyp}\begin{split}
	\mc Q_{\rsub{(e)}}[\dd\lambda; S] & = - \int_S\df{\dd \varepsilon}_2~ \dd\lambda \dd E_a \dd u^a \\
	\mc Q_{\rsub{(m)}}[\dd\lambda; S] & = \int_S\df{\dd \varepsilon}_2~ \dd\lambda \dd B_a \dd u^a 
\end{split}\ee
where \(\df{\dd\varepsilon}_2 \equiv \dd\varepsilon_{ab}\) is the area element on \(S\) (\cref{eq:volume-hyp}). Using the Maxwell equations \cref{eq:Max-hyp} the flux of these charges between any region \(\Delta\hyp\) of \(\hyp\) bounded by the cross-sections \(S_2\) and \(S_1\) (where \(S_2\) is in the future of \(S_1\)) is (with \(\df{\dd\varepsilon}_3 \equiv \dd\varepsilon_{abc}\) is the volume element on \(\hyp\))
\be\label{eq:flux-hyp}\begin{split}
    \mc F_{\rsub{(e)}}[\dd\lambda; \Delta\hyp] &\defn \mc Q_{\rsub{(e)}}[\dd\lambda; S_2] - \mc Q_{\rsub{(e)}}[\dd\lambda; S_1] = - \int_{\Delta\hyp} \df{\dd\varepsilon}_3~ \dd D^a \dd\lambda \dd E_a \\
    \mc F_{\rsub{(m)}}[\dd\lambda; \Delta\hyp] &\defn \mc Q_{\rsub{(m)}}[\dd\lambda; S_2] - \mc Q_{\rsub{(m)}}[\dd\lambda; S_1] = \int_{\Delta\hyp} \df{\dd\varepsilon}_3~ \dd D^a \dd\lambda \dd B_a 
\end{split}\ee

If the symmetry \(\lambda\) is \(C^0\) in spatial directions at \(i^0\) then the induced symmetry \(\dd\lambda = \text{constant}\) on \(\hyp\), and the fluxes \cref{eq:flux-hyp} vanish identically across \emph{any} region \(\Delta\hyp\). Only in this case, the corresponding charges \cref{eq:charge-hyp} are independent of the choice of cross-section of \(\hyp\) and are well-defined at \(i^0\) (these are the usual Coulomb charges at \(i^0\); see \cite{AH}). The rest of the charges can only be associated to the blowup \(\hyp\) and not to the asymptotic boundary \(i^0\) itself, in contrast to the charges for ``angle-dependent'' symmetries on \(\scri\). These additional charges on \(\hyp\) will be useful when relating the charges on \(\scri^-\) to those on \(\scri^+\).\\

Thus, in this picture the asymptotic symmetry algebra including all the asymptotic regions is the direct sum
\be\label{eq:symm-total}
    \mf g = \mf g^- \oplus \mf g^0 \oplus \mf g^+ \cong C^\infty(\bb S^2) \oplus C^\infty(\hyp) \oplus C^\infty(\bb S^2)
\ee
The direct sum structure of \(\mf g\) arises, essentially, because the hyperboloid \(\hyp\) does not ``attach'' to null directions along \(\scri\) at \(i^0\). As a result one cannot demand any ``continuity'' between the fields as they approach \(i^0\) along null directions and spatial directions; thus, the symmetries and fields on \(\scri\) and on \(\hyp\) are ``independent''. In fact, one can find examples of Maxwell fields which satisfy the conditions \cref{eq:F-asymp-spatial} along spatial directions at \(i^0\) but do not extend smoothly to null infinity, i.e. do not satisfy \cref{eq:F-asymp-null} (see \cite{Herb-dd}). In the next section we construct a blowup \(\cyl\) of \(i^0\) which does attach to the null directions at \(i^0\) along \(\scri\) and will allow us to relate the symmetries and fields on \(\scri\) to those on \(\hyp\) by imposing suitable regularity conditions in both null and spatial directions at \(i^0\).

\section{The space \(\cyl\) of null and spatial directions at \(i^0\)}
\label{sec:null-dir}

As discussed above the framework in \cref{sec:AH} is not adequate to analyse the behaviour of fields in the null directions along \(\scri^\pm\) at \(i^0\). Firstly, from \cref{cond:Omega-at-i0}, \(n^a\vert_{i^0} = 0\) and hence \(n^a\) does not specify any ``good'' null directions at \(i^0\). Secondly, on the hyperboloid \(\hyp\), null directions would (roughly speaking) correspond to points in the infinite future or past. In this section, we augment the Ashtekar-Hansen framework (still using \cref{def:AH}) and construct a different blowup of \(i^0\) which includes both null and spatial directions.

\begin{figure}[h!]
	\centering
	\includegraphics[width=0.3\textwidth]{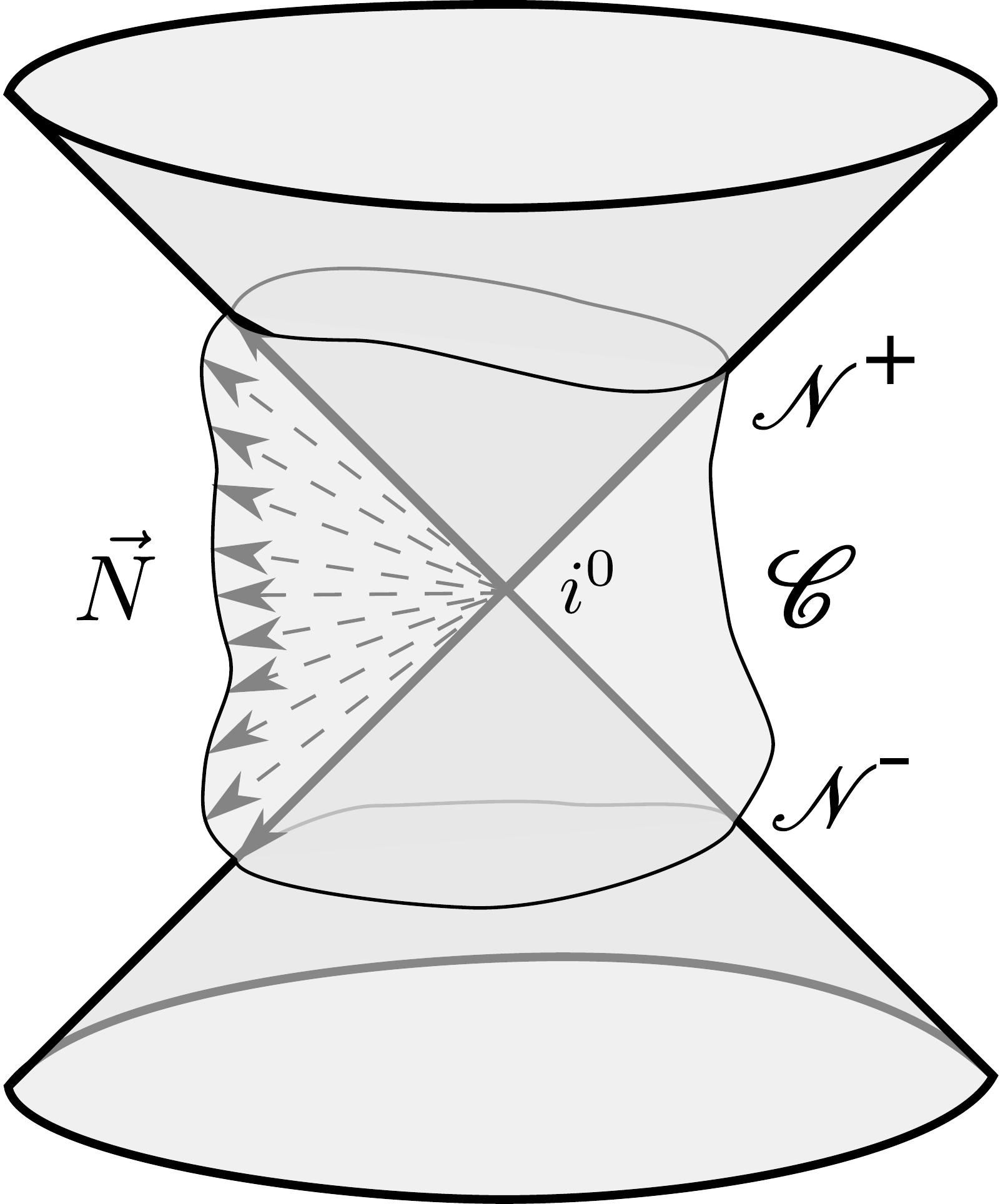}
	\caption{The finite cylinder \(\cyl\) in \(Ti^0\) representing both null and spatial directions \(\vec N\) at \(i^0\) (drawn as dashed rays). The boundaries \(\nulls^\pm \cong \bb S^2\) denote the space of null directions diffeomorphic to the space of generators of \(\scri^\pm\), respectively, while \(\cyl \setminus \nulls^\pm\) is the space of \emph{rescaled} spatial directions diffeomorphic to the unit-hyperboloid \(\hyp\). The space \(\cyl\) depends on the choice of the rescaling function \(\Sigma\) (defined below) and \emph{need not} be a cylinder of unit radius in \(Ti^0\) --- we have drawn a ``wiggly'' cylinder to emphasise this.} 
    \label{fig:cylinder}
\end{figure}

The above observations already suggest the following strategy.\footnote{This is similar to the strategy mentioned in the last footnote of \cite{Ash-in-Held}.}
\begin{enumerate*}
    \item We could rescale \(n^a\) so that the rescaled vector field is non-vanishing and represents ``good'' null directions at \(i^0\).
    \item We could also conformally-complete \(\hyp\) to get a new manifold whose boundaries represent the points in the infinite future or past along \(\hyp\).
    \item Then, if we can identify the null directions given by the rescaling of \(n^a\) with the boundaries of the conformal-completion of \(\hyp\) in a ``sufficiently smooth'' way, the new manifold will be the blowup we need.
\end{enumerate*}
In the rest of this section we will define a \emph{rescaling function} \(\Sigma\) which allows us to simultaneously implement all of the steps above, and enumerate its properties. In \cref{sec:Sigma}, we show that such a rescaling function does exist. We emphasise that the rescaling \(\Sigma\) is \emph{not} an alternative choice of the conformal factor \(\Omega\), in particular the unphysical metric \(g_{ab}\) is not rescaled by \(\Sigma\). The final picture we obtain is shown in \cref{fig:cylinder}.

In the following we will work in a small neighbourhood of \(i^0\) in \(M\), and use \(M\) to mean such a neighbourhood unless otherwise specified. In \(M\), we define a \emph{rescaling function} \(\Sigma\) as follows:

\begin{definition}[Rescaling function \(\Sigma\)]
\label{def:Sigma}
Let \(\Sigma\) be a function in \(M\) such that 
\begin{condlist}
    \item \(\Sigma^{-1} > 0\) is smooth on \(M - i^0\)
    \item \(\Sigma^{-1}\) is \(C^{>0}\) at \(i^0\) in both null and spatial directions,
	\item \(\Sigma^{-1}\vert_{i^0} = 0\), \(\lim_{\to i^0} \nabla_a \Sigma^{-1} \neq 0\) and \(\Sigma \Lie_n \Sigma^{-1}\vert_{i^0} = 2\) \label{cond:Sigma-i0}
\end{condlist}
\end{definition}
Note that the rescaling function \(\Sigma\), as defined above, can be chosen independently of the choice of conformal factor \(\Omega\).
\begin{remark}[Freedom in the rescaling function]
\label{rem:freedom-Sigma}
 The freedom in the choice of the rescaling function is given by \(\Sigma \mapsto \sigma \Sigma \) where the function \(\sigma\) satisfies
\begin{propertylist}
    \item \(\sigma > 0\) in \(M\)
    \item \(\sigma\) is smooth on \(M - i^0\)
    \item \(\sigma\) is \(C^{>-1}\) at \(i^0\) in both null and spatial directions
\end{propertylist}
\end{remark}
Using this freedom we choose a convenient rescaling function as follows. Let \(\Sigma' = \sigma\Sigma\), then
\be
    \Sigma' \Lie_n \Sigma'^{-1} = - \Lie_n \ln\sigma + \Sigma \Lie_n \Sigma^{-1}
\ee
Now since, \(\sigma > 0\) is \(C^{>-1}\) at \(i^0\) and \(n^a\vert_{i^0} = 0\) we have \(\Lie_n \sigma\vert_{i^0} = 0\). Thus we can always solve the equation
\be
    \Lie_n \ln\sigma\vert_\scri = (\Sigma\Lie_n \Sigma^{-1} - 2)\vert_\scri
\ee
so that \(\Sigma'\) satisfies \(\Sigma' \Lie_n \Sigma'^{-1} = 2\) not just at \(i^0\) but also on \(\scri\). Henceforth, we will restrict to rescaling functions satisfying
\be\label{eq:Sigma-choice}
    \Sigma \Lie_n \Sigma^{-1} = 2 \text{ at } i^0 \text{ and on } \scri
\ee 

\begin{remark}[Residual freedom in \(\Sigma\)]
\label{rem:freedom-Sigma-res}
    The choice \cref{eq:Sigma-choice}, now, depends on the choice of conformal factor \(\Omega\). It can be checked that if \(\Sigma\) is a choice of rescaling function for the conformal factor \(\Omega\) then \(\sigma\Sigma\) is a choice for the conformal factor \(\omega\Omega\) (both choices satisfying \cref{eq:Sigma-choice}) if
\be\label{eq:freedom-Sigma-res}
    \Lie_n \ln\sigma\vert_\scri = 2(1-\omega)\vert_\scri
\ee
where \(\omega\) satisfies the conditions in \cref{rem:freedom-Omega}.
\end{remark}

Any choice of the rescaling function \(\Sigma\) allows us to construct suitably regular fields near \(i^0\) which will be useful later. We list below their essential properties which can be verified as in \cref{sec:Sigma}.

Since \(\Sigma^{-1}\vert_{i^0} = 0\) and \(\Sigma^{-1}\) is \(C^{>0}\), there exists a function \(\dd\Sigma(\vec\eta)\), which is \(C^{>-1}\) along spatial directions, such that
\be\label{eq:dd-Sigma}
    \dd\Sigma^{-1}(\vec\eta) = \lim_{\to i^0} (\Omh \Sigma)^{-1}
\ee

\paragraph*{Rescaled normal \(N^a\) and the space \(\cyl\) of null and spatial directions:}
The rescaled vector field (the factor of half in definition of \(N^a\) is for later convenience)
\be\label{eq:N-defn} 
N^a \defn \tfrac{1}{2} \Sigma n^a = \tfrac{1}{2} \Sigma \nabla^a \Omega
\ee
is \(C^{>-1}\) at \(i^0\) such that \(\dd N^a = \lim_{\to i^0} N^a \neq 0\) in both null and spatial directions. Thus, along \(\scri\), we have \(\dd N^a\) as a direction-dependent null vector representing the null directions at \(i^0\) which are future/past directed along \(\scri^\pm\) respectively. Along spatial directions at \(i^0\) we have \(\dd N^a(\vec\eta) = \dd\Sigma(\vec\eta) \dd\eta^a \neq 0\) which represents the rescaled spatial directions at \(i^0\). The space of these directions \(\vec N\) can be represented by a cylinder \(\cyl\) with two boundaries \(\nulls^\pm \cong \bb S^2\) (as in \cref{fig:cylinder}). The boundaries \(\nulls^\pm\) represent the null directions along \(\scri^\pm\) respectively, while \(\cyl\setminus \nulls^\pm\) is the space of the rescaled spatial directions at \(i^0\).

\paragraph*{Rescaled auxilliary normal \(L^a\) and foliation of \(\scri\):}
Define a vector field \(L^a\) in \(M\) by
\be\label{eq:L-defn}
    L^a \defn - \nabla^a \Sigma^{-1} + \tfrac{1}{2}\nabla_b \Sigma^{-1} \nabla^b \Sigma^{-1} ~ N^a
\ee 
which is \(C^{>-1}\) at \(i^0\) and \(\lim_{\to i^0} L^a \neq 0\) in both null and spatial directions. Further, using \cref{eq:N-defn,eq:Sigma-choice}, we have
\be\label{eq:NL-LL}
    N^a L_a\vert_\scri = -1 \eqsp L^a L_a\vert_\scri = 0
\ee
The pullback to \(\scri\) of \(L_a\) equals the pullback of \(- \nabla_a \Sigma^{-1}\), thus \(L^a\) defines a rescaled auxilliary normal to the foliation of \(\scri\) by a family of cross-sections \(S_\Sigma\) with \(\Sigma^{-1} = \text{constant}\). From \cref{def:Sigma} and \cref{cond:int-curves}, the limiting cross-section \(S_\Sigma\) as \(\Sigma^{-1} \to 0\), is diffeomorphic to the space of null directions \(\nulls^\pm\). The auxilliary normal \(l^a\) to this foliation, satisfying \cref{eq:l-props}, is obtained by
\be\label{eq:l-defn}
    l^a \defn \tfrac{1}{2} \Sigma L^a
\ee
We also extend \(l^a\) into \(M\) by the above formula and \cref{eq:L-defn}.

\paragraph*{Conformal-completion of \(\hyp\):}
Let \(\dd\Sigma\) be the function induced on \(\hyp\) by \(\dd\Sigma(\vec\eta)\) (\cref{eq:dd-Sigma}). Let \((\tilde\hyp, \tilde{\dd h}_{ab}) \) be a conformal-completion of \((\hyp, \dd h_{ab})\) with metric  \(\tilde{\dd h}_{ab} = \dd\Sigma^2 \dd h_{ab}\). There exists a diffeomorphism from \(\tilde\hyp\) onto \(\cyl\) such that \(\hyp\) is mapped onto \(\cyl \setminus \nulls^\pm\) and \(\dd\Sigma\), as a function on \(\cyl \setminus \nulls^\pm\), extends smoothly to the boundaries \(\nulls^\pm\) where
\be
    \dd\Sigma\vert_{\nulls^\pm} = 0
\ee
Using \cref{eq:L-defn,eq:eta-defn,eq:d-eta-h,eq:dd-der-spatial}, the limit of \(L^a\) to \(i^0\) along spatial directions gives the direction-dependent vector field
\be
    \dd L^a(\vec\eta) = - \dd h^{ab} \dd D_b \dd\Sigma^{-1} + (\dd h^{bc} \dd D_b \dd\Sigma^{-1} \dd D_c \dd\Sigma^{-1}) \dd\eta^a
\ee
The projection of \(\dd L^a(\vec\eta)\) onto \(\hyp\) is the vector field
\be\label{eq:U-defn}
    \dd U^a \defn \dd h^a{}_b \dd L^b(\vec\eta) = \tilde{\dd h}^{ab} \dd D_b \dd \Sigma
\ee
Viewed as a vector field on \(\tilde\hyp\), and hence \(\cyl\), we have
\be\label{eq:U-on-N}
    \lim_{\to \nulls^\pm}\tilde{\dd h}_{ab} \dd U^a \dd U^b = -1 \eqsp \lim_{\to \nulls^\pm}\dd\Sigma^{-1} \dd U^a \neq 0 
\ee
Note that \(\dd\Sigma^{-1}\dd U^a\) is future/past directed at \(\nulls^\pm\) respectively (see \cref{tab:orientation}). Note, from \cref{eq:U-on-N}, the metric \(\tilde{\dd h}_{ab}\) is \emph{not} smooth at \(\nulls^\pm\) on \(\cyl\), but still provides a useful relation between \(\cyl\) and the conformal-completion of \(\hyp\).

\paragraph*{Metric on \(\nulls^\pm\):}
On \(\scri\) consider the rescaled metric
\be\label{eq:rescaled-q}
    \tilde q_{ab} \defn \Sigma^2 q_{ab}
\ee
Along the foliation \(S_\Sigma\) as \(\Sigma^{-1} \to 0\), \(\lim_{\to i^0} \tilde q_{ab}(\vec N)\) exists along null directions \(\vec N\) and defines a direction-dependent Riemannian metric \(\tilde{\dd q}_{ab}\) on the space of null directions \(\nulls^\pm\). Further, this metric coincides with the metric induced on \(\nulls^\pm\) by \(\tilde{\dd h}_{ab}\) on \(\cyl\), that is,
\be
    \tilde{\dd q}_{ab} = \lim_{\to \nulls^\pm} (\tilde{\dd h}_{ab} + \dd D_a \dd\Sigma \dd D_b \dd\Sigma)
\ee
Similarly, the rescaled area element \(\tilde\varepsilon_{ab} \defn \Sigma^2 \varepsilon_{ab}\) on the foliation \(S_\Sigma\) induces an area element \(\tilde{\dd\varepsilon}_{ab}\) on \(\nulls^\pm\) such that
\be\label{eq:nulls-area}
    \tilde{\dd\varepsilon}_{ab} = \lim_{\to \nulls^\pm} \dd U^c \tilde{\dd\varepsilon}_{cab} 
\ee
where \(\tilde{\dd\varepsilon}_{abc} \defn \dd\Sigma^3 \dd\varepsilon_{abc}\) is the volume element on \(\cyl\) defined by the conformal metric \(\tilde{\dd h}_{ab}\).

\paragraph*{Reflection conformal isometry of \(\cyl\):}
The reflection isometry \(\Upsilon\) of \(\hyp\) (see \cref{eq:reflection-hyp}) extends to a reflection conformal isometry of \(\cyl\) i.e., there exists a reflection map \(\Upsilon : \cyl \to \cyl\) and a smooth function \(\dd\varsigma > 0\) on \(\cyl\) such that
\be\label{eq:reflection-cyl}
    \Upsilon \circ \tilde{\dd h}_{ab} = \dd\varsigma^2 \tilde{\dd h}_{ab}
\ee
Further, under this map we also have \(\Upsilon : \nulls^- \to \nulls^+\) such that
\be\label{eq:reflection-cyl-boundary}\begin{split}
     \Upsilon \circ (\dd\Sigma^{-1} \dd U^a)\vert_{\nulls^-} & = - \dd\varsigma^{-2} \dd\Sigma^{-1} \dd U^a \vert_{\nulls^+} \\
    \Upsilon \circ \tilde{\dd \varepsilon}_{ab}\vert_{\nulls^-} & = - \dd\varsigma^2 \tilde{\dd \varepsilon}_{ab}\vert_{\nulls^+}
\end{split}\ee
where the negative signs on the right-hand-side are due to our orientation conventions (\cref{tab:orientation}).\\

If we choose a different rescaling function \(\Sigma' = \sigma \Sigma\) with \(\sigma\) satisfying the conditions in \cref{rem:freedom-Sigma,rem:freedom-Sigma-res}, we get a different space \(\cyl'\) of directions \(\vec N' = \dd\sigma \vec N\) at \(i^0\), where \(\dd\sigma = \lim_{\to i^0} \sigma\) along the directions \(\vec N\). This new space \(\cyl'\) is naturally diffeomorphic to \(\cyl\) under the above mapping of the directions with the metric \(\tilde{\dd h}_{ab}' = \dd \sigma^2 \tilde{\dd h}_{ab}\). Thus, we can treat \(\cyl\) as an abstract manifold with this conformal-class of metrics. The transformation of the various fields defined above can be computed directly from the defining equations.\\

The space \(\cyl\) enables us to impose continuity conditions on \(\nulls^\pm\) approaching \(i^0\) from both \(\scri^\pm\) and along spatial directions as follows. Let \(f\) be a function which is smooth at \(\scri\) and \(C^{>-1}\) at \(i^0\) in both null and spatial directions. Then along null directions, \(\lim_{\to i^0}f\vert_{\scri^\pm}\) induces a smooth function \(\dd f^\pm\) on \(\nulls^\pm\). Similarly, along spatial directions, \(\lim_{\to i^0} f \) induces a smooth function \(\dd f\) on \(\hyp\). Using the diffeomorphism between \(\hyp\) and \(\cyl \setminus \nulls^\pm\), we can consider \(\dd f\) as a smooth function on \(\cyl\setminus \nulls^\pm\). Since, \(f\) is \(C^{>-1}\) along both null and spatial directions, the function \(\dd f\) extends to \(\nulls^\pm\) as a smooth function, and on \(\nulls^\pm\) satisfies 
\be
    \dd f^\pm = \dd f \vert_{\nulls^\pm}
\ee
That is, for functions which are \(C^{>-1}\) in both null and spatial directions, the fields induced on \(\nulls^\pm\) by, first taking the limit to \(\scri^\pm\) and then to \(i^0\), or by, first taking the limit to \(i^0\) in spatial directions and then to the space of null directions \(\nulls^\pm\), coincide. Thus, such functions are continuous at the space of null directions \(\nulls^\pm\) when going from \(\cyl\) to \(\scri^\pm\).

\section{Null-regular Maxwell fields at \(i^0\)}
\label{sec:Maxwell-partial-matching}

With the blowup \(\cyl\), incorporating both null and spatial directions at \(i^0\), we can now impose suitable regularity conditions on the Maxwell fields on \(\cyl\) so that they smoothly extend to the space of null directions \(\nulls^\pm\) and ``match'' the fields induced on \(\nulls^\pm\) from \(\scri\). From the preceding discussion we see that functions which are \(C^{>-1}\) in both null and spatial directions satisfy this requirement. Hence, we now impose the following restriction on the Maxwell fields we consider.\footnote{Note, that we could have demanded that \(\Sigma^{-2}F_{ab}\) is a \(C^{>-1}\) tensor at \(i^0\), similar to the condition used by Herberthson \cite{Herb-dd}, but the weaker conditions in \cref{def:F-null-regular} will suffice for our purposes.} 

\begin{definition}[Null-regular Maxwell field at \(i^0\)]
\label{def:F-null-regular}
Let \(l^a\) be the vector field defined by \cref{eq:l-defn} in \(M\). We call a solution \(F_{ab}\) to Maxwell equations \emph{null-regular} at \(i^0\) if the rescaled quantities \(\Sigma^{-2} F_{ab}l^a n^b \) and \(\Sigma^{-2} (*F)_{ab}l^a n^b\) are \(C^{>-1}\) in \emph{both} null and spatial directions at \(i^0\).
\end{definition}

As we show below (\cref{rem:finite-flux-scri}), the above null-regularity conditions ensure that in a physical scattering process the flux of charges through \(\scri\) (including the point \(i^0\)) is finite. Thus the conditions in \cref{def:F-null-regular} discard the solutions which have infinite flux through null infinity and are hence unphysical in a scattering process. The form of these conditions in terms of the Bondi-Sachs parameter on \(\scri\) is given in \cref{eq:F-null-regular-BS}, which are consistent with the conditions imposed on Minkowski spacetime in \cite{CE}. In the following, we will focus on proving the matching problem for the electric charge and the argument for the magnetic charge follows from a similar analysis.

Using \cref{eq:N-defn,eq:l-defn,eq:E-B-defn,eq:U-defn}, the limit of \(\Sigma^{-2}F_{ab}l^a n^b\) to \(i^0\) in spatial directions can be rewritten as
\be
    \lim_{\to i^0} \Sigma^{-2}F_{ab}l^a n^b = \lim_{\to i^0}\Sigma^{-2} F_{ab} L^a N^b =  (\dd \Sigma^{-1} \dd E_a \dd U^a) (\vec\eta)
\ee
Thus, \(\Sigma^{-2}F_{ab}l^a n^b\) induces the field \(\dd \Sigma^{-1} \dd E_a \dd U^a\) on \(\hyp\) and hence on \(\cyl \setminus \nulls^\pm\). As discussed above, since \(\Sigma^{-2}F_{ab}l^a n^b\) is \(C^{>-1}\) in both null and spatial directions, \(\dd \Sigma^{-1} \dd E_a \dd U^a\) induces a smooth function on \(\nulls^\pm\) and coincides with the field induced by \(\lim_{\to i^0}\Sigma^{-2}F_{ab}l^a n^b\) on \(\nulls^\pm\) from \(\scri^\pm\) that is,
\be\label{eq:E-field-match}
    \lim_{\to i^0} (\Sigma^{-2}F_{ab}l^a n^b) \text{ along } \scri^\pm = \lim_{\to \nulls^\pm}(\dd \Sigma^{-1} \dd E_a \dd U^a) \text{ along } \cyl
\ee\\

We can now ask about the limits of the charges \cref{eq:charge-null,eq:charge-hyp} as the cross-sections of \(\scri\) and \(\cyl\) tend to \(\nulls^\pm\). First, consider the electric charge (\cref{eq:charge-null}) associated to some symmetry \(\lambda^\pm \in \mf g^\pm\) evaluated on a cross-section \(S_\Sigma\) of \(\scri^\pm\) with \(\Sigma^{-1} = \text{constant}\). Since, \(\Lie_n \lambda^\pm = 0\), \(\lim_{\to i^0} \lambda^\pm\) exists and induces a smooth function \(\dd\lambda^\pm\) on \(\nulls^\pm\). Similarly, \(\lim_{\to i^0} \Sigma^2 \df\varepsilon_{ab}\) induces the area element \(\tilde{\df{\dd\varepsilon}}_{ab}\) on \(\nulls^\pm\). Thus, as the cross-section \(S_\Sigma\) tends to \(i^0\), i.e. \(\Sigma^{-1} \to 0\) the electric charge induced on \(\nulls^\pm\) from \(\scri^\pm\) is given by
\be\label{eq:charge-null-scri}
    \mc Q_{\rsub{(e)}}[\lambda^\pm; \nulls^\pm] = \lim_{\Sigma^{-1} \to 0}~ - \int_{S_\Sigma} \df\varepsilon_2~ \lambda F_{ab}l^a n^b = - \int_{\nulls^\pm} \tilde{\df{\dd \varepsilon}}_2~ \dd\lambda^\pm (\Sigma^{-2}F_{ab}l^a n^b) 
\ee
which is finite due to our null-regularity conditions.

\begin{remark}[Finiteness of flux through \(\scri\)]
\label{rem:finite-flux-scri}
Let \(\Delta \scri\) be a region of \(\scri\) foliated by cross-sections \(S_\Sigma\) with \(\Sigma^{-1}_0 \leq \Sigma^{-1} \leq \Sigma^{-1}_1\). Since the limit to \(i^0\) of the charge evaluated on \(S_\Sigma\) exists, the flux of charge must be finite as \(\Sigma^{-1}_0 \to 0\). However, the volume element \(\varepsilon_{abc}\) on \(\scri\) (appearing in \cref{eq:flux-null}) is ill-defined in this limit since \(l^a\) diverges at \(i^0\). But the rescaled volume element
\be
    L^d \varepsilon_{dabc} = 2 \Sigma^{-1} \varepsilon_{abc} \text{ is } C^{>-1} \text{ at } i^0
\ee
Further, since the pullback of \(L_a\) to \(\scri\) is the pullback of \(-\nabla_a \Sigma^{-1}\) we also have
\be
    L^d \varepsilon_{dabc} = 3 \nabla_{[a} \Sigma^{-1} \varepsilon_{bc]}
\ee
Thus, on \(\Delta \scri\) foliated by \(S_\Sigma\) we can rewrite the flux \cref{eq:flux-null} as 
\be\begin{split}
    \mc F_{\rsub{(e)}}[\lambda; \Delta\scri] 
    & = \int_{\Sigma^{-1}_0}^{\Sigma^{-1}_1} d \Sigma^{-1}~ \Sigma \int_{S_\Sigma} \tilde{\df\varepsilon}_2~ \tfrac{1}{2} \lb(\lambda \Sigma^{-2}n_a J^a + \tilde q^{ab} \nabla_a\lambda \mc E_b \rb)
\end{split}\ee
where we have used the rescaled metric \(\tilde q^{ab} = \Sigma^{-2} q^{ab}\) and area element \(\tilde\varepsilon_{ab} = \Sigma^2 \varepsilon_{ab}\) on the cross-sections \(S_\Sigma\). Since in the limit \(\Sigma^{-1}_0 \to 0\), the flux on the left-hand-side is finite, and \(\tilde q^{ab}\), \(\tilde{\df\varepsilon}_2\) and \(\lambda\) on the right-hand-side induce smooth fields on \(\nulls^\pm\), we get the falloffs 
\be\label{eq:J-E-null-regular}
    \Sigma^{-2} n_a J^a = O(\Sigma^{-\epsilon}) \eqsp \mc E_a = O(\Sigma^{-\epsilon}) \quad\text{along null directions}
\ee
for some small \(\epsilon > 0\). In terms of the Bondi-Sachs parameter these falloffs are given in \cref{eq:J-E-null-regular-BS}.
\end{remark}

Now we evaluate the charge \cref{eq:charge-hyp} on \(\cyl\) as the cross-section \(S \to \nulls^\pm\). Since, the normals \(\dd u^a\) and \(\dd U^a\) are timelike and unit with respect to the metrics \(\dd h_{ab}\) and \(\tilde{\dd h}_{ab}\), respectively, we have
\be
    \lim_{\to \nulls^\pm} \dd\Sigma^{-1}\dd U^a = \pm \lim_{\to \nulls^\pm} \dd\Sigma^{-2} \dd u^a
\ee
where the signs on the right-hand-side are due to our orientation conventions from \cref{tab:orientation}. Similarly, we have for the area element on \(S\) in the limit to \(\nulls^\pm\) (from \cref{eq:volume-hyp,eq:nulls-area})
\be
    \tilde{\dd\varepsilon}_{ab} = \pm \lim_{S \to \nulls^\pm} \dd\Sigma^2 \dd\varepsilon_{ab} 
\ee
The symmetry \(\dd \lambda \in \mf g^0\) is a smooth function on \(\hyp\) and hence on \(\cyl\setminus\nulls^\pm\). However, \(\dd\lambda\) need not have a limit to \(\nulls^\pm\), and the limit of the charge in \cref{eq:charge-hyp} need not exist even if the corresponding fields are smooth. Thus, we now restrict to the symmetries \(\dd\lambda\) which extend smoothly to \(\nulls^\pm\). Then the charge induced on \(\nulls^\pm\) from \(\cyl\) is finite and is given by
\be\label{eq:charge-null-C}
    \mc Q_{\rsub{(e)}}[\dd\lambda; \nulls^\pm] = - \lim_{S \to \nulls^\pm} \int_S \df{\dd \varepsilon}_2~ \dd\lambda \dd E_a \dd u^a = - \int_{\nulls^\pm} \tilde{\df{\dd \varepsilon}}_2~ \dd\lambda (\dd\Sigma^{-1}\dd E_a \dd U^a)
\ee

A priori, the charges induced on \(\nulls^\pm\) by \cref{eq:charge-null-scri,eq:charge-null-C} need not be equal, even accounting for \cref{eq:E-field-match}, since the symmetry can have a discontinuous ``jump'' at \(\nulls^\pm\) from \(\dd\lambda\) on \(\cyl\) to \(\lambda^\pm\) on \(\scri^\pm\). Thus, for null-regular Maxwell fields, we consider the subalgebra \(\mf g^{\rsub{nr}} \subset \mf g\) ---  which we call the \emph{null-regular} symmetry algebra --- given by
\be
    \mf g^{\rsub{nr}} = \set{\dd\lambda \st (\lambda^-, \dd\lambda, \lambda^+) \in \mf g \text{ and } \lim_{\to i^0} \lambda^\pm \text{ along }\scri^\pm = \lim_{\to \nulls^\pm} \dd\lambda \text{ along }\cyl }
\ee 
Thus, the elements of the reduced symmetry algebra are completely determined by the function \(\dd\lambda\) on \(\cyl\) and the symmetries on \(\scri^\pm\)  are given by the boundary values of \(\dd\lambda\) on \(\nulls^\pm\), respectively. That is, we have the following Lie algebra homomorphism from \(\mf g^{\rsub{nr}}\) to each of \(\mf g^\pm\)
\be
    \mf g^{\rsub{nr}} \to \mf g^\pm \st \dd\lambda \mapsto \lambda^\pm = \dd\lambda\vert_{\nulls^\pm}
\ee
which can be considered as a partial matching of the symmetries at \(i^0\) to those on \(\scri^\pm\). For this subalgebra, from \cref{eq:E-field-match,eq:charge-null-scri,eq:charge-null-C} we have the partial matching of charges at \(\nulls^\pm\),
\be\label{eq:charge-partial-matching}
    \mc Q_{\rsub{(e)}}[\lambda^\pm; \nulls^\pm] \text{ along } \scri^\pm = \mc Q_{\rsub{(e)}}[\dd\lambda; \nulls^\pm] \text{ along } \cyl
\ee

The null-regular symmetries \(\mf g^{\rsub{nr}}\) do not provide a unique isomorphism from \(\mf g^-\) to \(\mf g^+\). From a given symmetry \(\lambda^-\) on \(\scri^-\) we can get any symmetry \(\lambda^+\) on \(\scri^+\) simply by choosing \(\dd\lambda\) suitably on \(\cyl\). We show next that a \emph{natural} choice for a subalgebra of \(\mf g^{\rsub{nr}}\) exists, which is suitable for scattering problems and provides a natural isomorphism from \(\mf g^-\) to \(\mf g^+\).

\subsection{Totally fluxless symmetries on \(\cyl\), antipodal matching of symmetries, and conservation laws}
\label{sec:Maxwell-matching}

For some arbitrary choice of \(\dd\lambda \in \mf g^{\rsub{nr}}\), the flux of charge through all of \(\cyl\) need not vanish even for the electric field generated by a static charge on Minkowski spacetime. To see this, the flux through all of \(\cyl\) is given by the difference of the charge integrals on \(\nulls^\pm\) i.e.,
\be
    \mc F_{\rsub{(e)}}[\dd\lambda; \cyl] = \int_{\nulls^+} \df{\tilde{\dd \varepsilon}}_2~ \dd\lambda^+ (\dd\Sigma^{-1} \dd E_a \dd U^a) - \int_{\nulls^-} \df{\tilde{\dd \varepsilon}}_2~ \dd\lambda^- (\dd\Sigma^{-1} \dd E_a \dd U^a)
\ee
where \(\dd\lambda^\pm = \dd\lambda \vert_{\nulls^\pm}\). From the analysis in \cref{sec:wave-hyp}, the only solutions to Maxwell equations on \(\cyl\) for which \(\lim_{\to \nulls^\pm}\dd\Sigma^{-1} \dd E_a \dd U^a\) exists (and thus corresponds to null-regular Maxwell fields) are the ones which satisfy
\be\label{eq:reflection-electric}
    \Upsilon \circ (\dd\Sigma^{-1} \dd E_a \dd U^a)\vert_{\nulls^-} = - (\dd\varsigma^{-2} \dd\Sigma^{-1} \dd E_a \dd U^a)\vert_{\nulls^+}
\ee
where \(\Upsilon\) is the reflection conformal isometry of \(\cyl\) (\cref{eq:reflection-cyl}). Then, using the transformation of \(\tilde{\dd\varepsilon}_{ab}\) under \(\Upsilon\), \cref{eq:reflection-cyl-boundary}, the total flux is given by
\be\label{eq:tot-flux-cyl}
    \mc F_{\rsub{(e)}}[\dd\lambda; \cyl] = \int_{\nulls^+} \df{\tilde{\dd \varepsilon}}_2~ (\dd\lambda^+ - \Upsilon \circ \dd\lambda^-) (\dd\Sigma^{-1} \dd E_a \dd U^a)
\ee
Since, \((\dd\lambda^+ - \Upsilon \circ \dd\lambda^-)\) is any function on \(\bb S^2\), the total flux through \(\cyl\) can take any value even for the pure Coulomb field of a static charge (see \cref{sec:wave-hyp}). This suggests that the total flux through \(\cyl\) is ``spurious'' as is to be expected in any scattering process. Thus, for scattering problems we further restrict the symmetry algebra to those elements which have vanishing total flux on \(\cyl\). From \cref{eq:tot-flux-cyl}, the only symmetries \(\dd\lambda\) which satisfy \(\mc F_{\rsub{(e)}}[\dd\lambda; \cyl] = 0\) are the ones for which
\be\label{eq:symm-antipodal}
    \Upsilon \circ \dd\lambda^- = \dd\lambda^+ 
\ee
Note that this is not a restriction on the Maxwell fields unlike the null-regularity conditions in \cref{def:F-null-regular}. The behaviour of \(\dd\lambda\) in \(\cyl\setminus\nulls^\pm\) can be arbitrary, and only the boundary values at \(\nulls^\pm\) are required to satisfy \cref{eq:symm-antipodal}. Thus, we define the equivalence class \([\dd\lambda]\) for any \(\dd\lambda\) satisfying \cref{eq:symm-antipodal} by
\be\label{eq:equiv-fluxless-symm}
    \dd\lambda' \in [\dd\lambda] \iff \dd\lambda'\vert_{\nulls\pm} = \dd\lambda\vert_{\nulls\pm} 
\ee
Each equivalence class \([\dd\lambda]\) is uniquely determined by a smooth function on \(\bb S^2\), either considered as a function on \(\nulls^-\) or \(\nulls^+\) related by \cref{eq:symm-antipodal}. Thus, the condition that the total flux through \(\cyl\) vanish gives us the diagonal symmetry algebra
\be\label{eq:tot-fluxless-symm}
    \mf g^\times \defn \set{ [\dd\lambda] \st \dd\lambda \in \mf g^{\rsub{nr}} \text{ and } \Upsilon \circ \dd\lambda^- = \dd\lambda^+  }  \cong C^\infty(\bb S^2)
\ee
The symmetries in \(\mf g^\times\) provide a natural isomorphism between the asymptotic symmetries \(\mf g^-\) and \(\mf g^+\) on null infinity as follows. Any symmetry \(\lambda^-\) on \(\scri^-\) determines a unique \([\dd\lambda] \in \mf g^\times\) on \(\cyl\) so that \(\lambda^- = \dd\lambda\vert_{\nulls^-}\). From \cref{eq:tot-fluxless-symm} this determines a unique symmetry \(\lambda^+\) on \(\scri^+\) by \(\lambda^+ = \dd\lambda\vert_{\nulls^+}\), and we have the isomorphism
\be\label{eq:symm-match}
    \mf g^- \to \mf g^+ : \lambda^-(\theta^A) \mapsto \lambda^+(-\theta^A)
\ee
That is, for the subalgebra \(\mf g^\times\), the symmetries on \(\scri^-\) can be matched to those on \(\scri^+\) through an antipodal reflection on \(\bb S^2\). From \cref{eq:charge-partial-matching}, we see that under the isomorphism \cref{eq:symm-match} we have
\be\label{eq:charge-matching}
    \mc Q_{\rsub{(e)}}[\lambda^-; \nulls^-] \text{ along } \scri^- = \mc Q_{\rsub{(e)}}[\lambda^+; \nulls^+] \text{ along } \scri^+
\ee
as a direct consequence of the corresponding symmetry \(\dd\lambda\) on \(\cyl\) being totally fluxless. This resolves the matching problem, as conjectured by Strominger \cite{Stro-CK-match}, for the asymptotic symmetries for null-regular Maxwell fields.

\begin{remark}[Change of rescaling function and conformal factor]
\label{rem:change-Sigma}
From \cref{rem:freedom-Omega}, the freedom \(\omega\) in the conformal factor satisfies \(\omega\vert_{i^0} = 1\) and hence our analysis is independent of the choice of conformal factor. Thus it suffices to consider the change of the rescaling function \(\Sigma \mapsto \sigma \Sigma\) (\cref{rem:freedom-Sigma,rem:freedom-Sigma-res}) where \(\Lie_n \sigma = 0\) on \(\scri\). Note that our null-regularity conditions \cref{def:F-null-regular} are independent of \(\sigma\) since \(\sigma\) is \(C^{>-1}\) at \(i^0\). Then, using \(\Sigma^{-2} F_{ab}l^a n^b = \Sigma^{-2} F_{ab}L^a N^b \), \cref{eq:N-defn,eq:L-defn}, we can compute\footnote{The transformation of \(L^a\) under the change of rescaling function also includes terms proportional to \(N^a\) but these drop out of \cref{eq:change-Fln} since \(F_{ab}\) is antisymmetric.}
\be\label{eq:change-Fln}\begin{split}
    \Sigma^{-2} F_{ab}l^a n^b\vert_\scri & \mapsto \sigma^{-2} \lb[ (\Sigma^{-2} F_{ab}l^a n^b)  - \tfrac{1}{2} \Lie_n \ln\sigma (\Sigma^{-2}F_{ab}l^a n^b) + \tfrac{1}{2} \Sigma^{-2} q^{ab} \nabla_a \ln\sigma \mc E_b \rb]\vert_\scri \\
    & = \sigma^{-2} \lb[\Sigma^{-2} F_{ab}l^a n^b + \tfrac{1}{2} \tilde q^{ab} \nabla_a \ln\sigma \mc E_b \rb]\vert_\scri
\end{split}\ee
where, in the second line, we have used \(\Lie_n \sigma = 0\) and converted to \(\tilde q^{ab} = \Sigma^{-2} q^{ab}\). Since \(\Sigma^{-2} F_{ab}l^a n^b\) and \(\sigma > 0\) are \(C^{>-1}\) in null directions, \(\lim_{\to i^0}\tilde q^{ab}\) is a non-degenerate metric on \(\bb S^2\), and from the falloff in \cref{eq:J-E-null-regular} we get in the limit to \(i^0\)
\be
    (\Sigma^{-2}F_{ab}l^a n^b) \vert_{\nulls^\pm} \mapsto (\dd\sigma^\pm)^{-2} (\Sigma^{-2}F_{ab}l^a n^b) \vert_{\nulls^\pm}
\ee
where \(\dd\sigma^\pm\) is the function induced on \(\nulls^\pm\) by \(\lim_{\to i^0} \sigma\) along \(\scri^\pm\).
Similarly on \(\cyl\), using \cref{eq:U-defn}, we have
\be\begin{split}
    (\dd \Sigma^{-1} \dd E_a \dd U^a)\vert_{\nulls^\pm} \mapsto (\dd\sigma^\pm)^{-2} (\dd \Sigma^{-1} \dd E_a \dd U^a)\vert_{\nulls^\pm}
\end{split}\ee
where we have used the fact that \(\sigma\) is \(C^{>-1}\) in both null and spatial directions so that on \(\cyl\), \(\lim_{\to \nulls^\pm} \dd \sigma = \dd\sigma^\pm\).
Further, the area element \(\tilde{\dd \varepsilon}_{ab}\) induced on \(\nulls^\pm\) from both \(\scri\) and \(\cyl\) transforms as
\be
    \tilde{\dd \varepsilon}_{ab} \mapsto (\dd \sigma^\pm)^2 \tilde{\dd \varepsilon}_{ab}
\ee
Thus, the charges induced on \(\nulls^\pm\) from both \(\scri\) and \(\cyl\), and our matching result are independent of the choice of rescaling function and conformal factor. Using the analysis of \cite{Ash-log}, it can also be checked that our result is unaffected by the logarithmic translation ambiguity in the choice of the Ashtekar-Hansen structure.
\end{remark}

\begin{remark}[Gauge choice in the physical spacetime]
The above matching result was also obtained by \cite{CE} on a Minkowski background where they imposed the Lorenz gauge on the Maxwell vector potentials in the physical spacetime. On \(\hyp\), the Lorenz gauge restricts the asymptotic symmetries (i.e. ``residual'' gauge freedom) to satisfy the wave equation \(\dd D^a \dd D_a \dd\lambda = 0\), which can be solved as in \cref{sec:wave-hyp}. This wave equation on \(\hyp\) was also obtained by \cite{HT-em} by adding certain boundary terms to the Maxwell action. Then, \cite{CE, HT-em} use the solutions for \(\dd\lambda\) corresponding to \(\dd V^{\rsub{(even)}}\) in \cref{eq:wave-soln} which do satisfy \cref{eq:symm-antipodal}, and hence determine an element of the diagonal symmetry algebra \(\mf g^\times\). By contrast, our analysis is done in a gauge-invariant manner purely on the asymptotic boundaries.
\end{remark}

\hr

With the diagonal symmetry algebra \(\mf g^\times\) we can now analyse the conservation of flux between \(\scri^-\) and \(\scri^+\). Consider any symmetry \([\dd\lambda] \in \mf g^\times\), and let \(\lambda^\pm \in \mf g^\pm\) be the unique symmetries on \(\scri^\pm\) determined by the boundary values on \(\nulls^\pm\) of any representative \(\dd \lambda \in [\dd\lambda]\). Let \(S^\pm\) be some (finite) cross-sections of \(\scri^\pm\), respectively, and let \(\mc F^\pm\) denote the (electric or magnetic) flux between \(i^0\) and \(S^\pm\) corresponding to \(\lambda^\pm\). Note that in our convention both \(\mc F^\pm\) are \emph{incoming} fluxes into the physical spacetime.

From the preceding analysis we know that for any symmetry \([\dd\lambda] \in \mf g^\times\) the corresponding charges for \(\lambda^\pm\) at \(i^0\) from both \(\scri^\pm\) match (\cref{eq:charge-matching}), and the fluxes \(\mc F^\pm\) are finite (\cref{rem:finite-flux-scri}). This immediately gives us the conservation law for charges defined on \(S^\pm\)
\be\label{eq:local-conservation}
    \mc Q[\lambda^+; S^+] - \mc Q[\lambda^-; S^-] = \mc F^+[\lambda^+] + \mc F^-[\lambda^-]
\ee 

Further, if \(\scri^\pm\) each have future/past boundaries at timelike infinities \(i^\pm\), respectively, and the Maxwell fields satisfy appropriate conditions at \(i^\pm\) (see \cref{rem:timelike-infinity}) so that the charges vanish as \(S^\pm \to i^\pm\), then we have the \emph{global} conservation law
\be\label{eq:global-conservation}
    \mc F^+[\lambda^+; \scri^+] + \mc F^-[\lambda^-; \scri^-] = 0
\ee
This implies that the total \emph{incoming} flux on \(\scri^-\) equals the total \emph{outgoing} flux at \(\scri^+\) and thus the flux is conserved in the scattering process from \(\scri^-\) to \(\scri^+\).

\begin{remark}[Timelike infinities \(i^\pm\)]
\label{rem:timelike-infinity}
To derive the global conservation law \cref{eq:global-conservation}, one needs to impose suitable falloff conditions on the Maxwell fields and sources at timelike infinities \(i^\pm\). However these falloff conditions cannot be chosen freely. Unlike at \(i^0\), the behaviour of the fields at \(i^\pm\) is completely determined by the equations of motion along with suitable initial data. Any falloff conditions one imposes at \(i^\pm\) should allow for, at the very least, solutions to Maxwell equations with initial data which is compactly-supported on a Cauchy surface, or more generally, initial data which is asymptotically-flat at \(i^0\) in the sense of \cref{eq:F-asymp-spatial}. Then, if the Maxwell field satisfies suitable falloffs approaching \(i^\pm\), analogous to the conditions at \(i^0\), one can adapt the Ashtekar-Hansen formalism and our method above to derive global conservation laws as in \cref{eq:global-conservation}; see \cite{Porrill, Cutler} for the cases where the spacetime becomes asymptotically-flat at \(i^\pm\). In these cases, for each \(i^\pm\), the analogues of the blowups \(\hyp\) and \(\cyl\) are a spacelike-unit-hyperboloid and a \(3\)-dimensional ball with a single boundary, respectively. The situation for spacetimes which contain black holes is much more complicated. The presence of an event horizon implies that the metric \(g_{ab}\) \emph{cannot} be continuous at \(i^\pm\) and can only be assumed to be \(C^{>-1}\). In this case, the Ashtekar-Hansen formalism and our method cannot be used directly, and a more careful treatment is needed \cite{HL-timelike}. Further, one would have to account for the charges associated to suitably defined symmetries on the event horizon (see, for instance, \cite{CFP}).
\end{remark}

\section{Discussion and possible generalisations}
\label{sec:disc}

We showed that suitably regular Maxwell fields satisfy the matching conditions for asymptotic symmetries and charges conjectured in \cite{Stro-CK-match} in any asymptotically-flat spacetime. The key steps in our method are
\begin{enumerate*}
    \item the construction of the space \(\cyl\) which allows us to simultaneously consider the limits of (suitably rescaled) Maxwell fields in both null and spatial directions at \(i^0\). 
    \item the null-regularity conditions on the Maxwell fields along null directions in \cref{def:F-null-regular} which ensure that the charges defined on \(\scri\) admit limits as one approaches \(i^0\) and these limits match the charges on \(\cyl\).
    \item the choice of the totally fluxless subalgebra of symmetries on \(\cyl\), reflecting the physical criteria that there is no flux across spatial infinity in a scattering process.
\end{enumerate*}
These, along with the Maxwell equations on \(\cyl\) near spatial infinity, then imply that the fields from \(\scri^-\) to \(\scri^+\) antipodally match at \(i^0\) and reduce the symmetry algebra to the diagonal subalgebra \(\mf g^\times\). As a consequence we showed that the charges associated to \(\mf g^\times\) on \(\scri^\pm\) match at \(i^0\) and the associated fluxes are conserved.

Note that the null-regularity conditions (\cref{def:F-null-regular}) imposed on the Maxwell fields are crucial in our analysis, as they ensure that the flux of charges radiated through \(\scri^\pm\) in the scattering process are finite (\cref{rem:finite-flux-scri}). Solutions to the Maxwell equations which do not satisfy our null-regularity conditions can be easily constructed --- see \cite{Herb-dd} for an example on Minkowski spacetime, and \cref{sec:wave-hyp} for the asymptotic solutions near \(i^0\) (with a reflection-even potential \(\dd V = \dd V^{\rsub{(even)}}\) on \(\hyp\)) in any spacetime with an Ashtekar-Hansen structure. However, for such solutions the flux of charges radiated through null infinity will not be finite, and so such solutions are unphysical from the point of view of a scattering problem. If one includes such solutions, one can still define a diagonal symmetry algebra \(\mf g^\times\) by simply demanding that the antipodal matching condition \cref{eq:symm-match} holds. But even with this assumed matching condition on the symmetries, one would not have any corresponding conservation laws (similar to \cref{eq:local-conservation,eq:global-conservation}) since the fluxes diverge.

 We also emphasise that our analysis is both gauge-invariant and spacetime-covariant, i.e., we do not use any gauge conditions on the fields or a \((3+1)\)-decomposition in the physical spacetime. We expect that our methods can be used to prove similar matching conditions for \(p\)-form gauge fields \cite{AES-p-form}, Yang-Mills theory \cite{Stro-YM}, and linearised gravity \cite{Tro} on asymptotically-flat spacetimes with suitable regularity conditions on the fields analogous to \cref{def:F-null-regular}. \\

The situation in full non-perturbative general relativity is as follows. Ashtekar and Magnon-Ashtekar \cite{Ash-Mag-Ash} showed that the limit of Bondi energy-momentum to \(i^0\) along both \(\scri^\pm\) equals the ADM energy-momentum. Similar result for the Bondi and ADM angular momentum was obtained by Ashtekar and Streubel \cite{AS-ang-mom} for spacetimes which become stationary ``sufficiently fast'' near \(i^0\). The antipodal matching for general supertranslations and charges in the class of CK-spacetimes was shown by Strominger \cite{Stro-CK-match}. However, the charges for all the angle-dependent supertranslations vanish in the limit to \(i^0\) along \(\scri\) in CK-spacetimes \cite{Ash-CK-triv}, and only the Bondi energy-momentum associated to translations is non-vanishing for which the matching problem was already solved in \cite{Ash-Mag-Ash}. Thus, even though the class of CK-spacetimes forms an open ball (in some suitable topology) around Minkowski spacetime, it is not general enough to address the non-trivial aspects of the matching problem for supertranslations.

We expect that our method can be applied to supertranslation symmetries and charges in general relativity with some modifications as follows. In the analysis of this paper we only assumed that the unphysical metric \(g_{ab}\) is \(C^0\) at \(i^0\) along null directions and, in general, the metric connection need not have a limit to \(i^0\) in null directions. This is sufficient for Maxwell fields since the Maxwell equations and charges are independent of the metric connection. However, in general relativity the charges for supertranslations on \(\scri\) depend on, not only the asymptotic Weyl curvature, but also the News tensor and some connection coefficients (or derivatives of the null frames) \cite{GW, AS-symp, WZ}. Thus, for supertranslations in general relativity one would have to assume additional regularity conditions on the metric along null directions to ensure that the charges and fluxes along \(\scri\) remain finite as one approaches \(i^0\). Such conditions were already necessary in \cite{Ash-Mag-Ash} for the case of the Bondi energy-momentum. One could simply assume that \(g_{ab}\) is \(C^{>0}\) in both null and spatial directions, but it is unclear if this assumption is too strong and rules out physically interesting spacetimes \cite{Ash-Mag-Ash, HL-GR-matching}. However from the Maxwell case considered here, we see that one only needs such regularity conditions on \emph{some} components of the relevant fields. A non-trivial result in this direction was obtained by Herberthson and Ludvigsen \cite{HL-GR-matching} who proved that the leading-order Weyl tensor component \(\Psi_2^0\) on \(\scri^-\) matches antipodally with the corresponding quantity on \(\scri^+\). This earlier result resolves the matching problem for a class of spacetimes --- much more general than the later work by Strominger \cite{Stro-CK-match} --- where the News tensor and connection components falloff fast enough in the limit to \(i^0\). With suitable generalisations of the null-regularity conditions and totally fluxless symmetry algebra at \(i^0\) from \cref{sec:Maxwell-partial-matching}, it should be possible to prove the matching of supertranslation symmetries and charges in general relativity in a large class of spacetimes. We defer the detailed investigation of this problem to future work.

The analogous analysis for the charges associated to the full BMS algebra is trickier. In this case, we run into the well-known supertranslation ambiguities in defining the charges for a Lorentz subalgebra, such as angular momentum, on null infinity. For stationary spacetimes we can define angular momentum unambiguously at the cost of reducing the BMS algebra to the Poincar\'e algebra \cite{NP-red}. To define angular momentum at spatial infinity, we either have to impose the Regge-Teitelboim  parity conditions \cite{RT-parity} in a \((3+1)\)-formalism, or, stronger falloff conditions on the magnetic part of the Weyl tensor at \(i^0\) in the Ashtekar-Hansen framework \cite{AH}. Both of these reduce the asymptotic symmetry algebra at \(i^0\) to the Poincar\'e algebra and we again lose access to the supertranslations. In this symmetry-reduced case, the matching conditions for angular momentum (and other Lorentz charges) follow from \cite{AS-ang-mom}. But, any attempt to prove the matching conditions for the entire BMS algebra of symmetries and their charges, would have to resolve the dichotomy between having well-defined Lorentz charges and including supertranslations as asymptotic symmetries (see, for instance, \cite{HT} where an alternative set of parity conditions were proposed in a \((3+1)\)-formalism).

\section*{Acknowledgements}
I would like to thank \'Eanna \'E. Flanagan for suggesting this problem, and constant encouragement and discussions throughout this work. I would also like to thank Mr.~Mark for inspiration. This work is supported in part by the NSF grants PHY-1404105 and PHY-1707800 to Cornell University.

\appendix

\section{\(C^{>1}\) differential structure and direction-dependent tensors}
\label{sec:dd}

Consider a manifold \(M\) which is smooth everywhere except at a point \(p \in M\) where it is \(C^1\) --- so that the tangent space \(Tp\) at \(p\) is well-defined. A function \(f\) in \(M\) is \emph{direction-dependent} at \(p\) if the limit of \(f\) along any curve \(\Gamma\), which is \(C^1\) at \(p\), exists and depends only on the tangent direction to \(\Gamma\) at \(p\). We write this as \(\lim_{\to p} f = \dd f(\vec N)\) where \(\vec N\) is the direction of the tangent to \(\Gamma\) at \(p\). Note that we can consider the limit \(\dd f(\vec N)\) as a function in the tangent space \(Tp\) which is constant along the rays (represented by \(\vec N\)) from \(p\).

\begin{definition}[Regular direction-dependent function \cite{Herb-dd}]
Let \(x^i\) denote a \(C^1\) coordinate chart with \(x^i(p) = 0\). A direction-dependent function \(f\) is \emph{regular direction-dependent} at \(p\) (with respect to the chosen chart \(x^i\)) if for all \(k\)
\be\label{eq:dd-derivative}
    \lim_{\to p} \lb( x^{i_1} \frac{\partial}{\partial x^{i_1}} \rb) \cdots \lb( x^{i_k} \frac{\partial}{\partial x^{i_k}} \rb) f = \lb[ \lb( x^{i_1} \frac{\partial}{\partial x^{i_1}} \rb) \cdots \lb( x^{i_k} \frac{\partial}{\partial x^{i_k}} \rb) \dd f \rb] (\vec N)
\ee
where on the right-hand-side we consider \(\dd f\) as a function in \(Tp\) as mentioned above. 
\end{definition}
\cref{eq:dd-derivative} ensures that, in the limit, \(f\) is smooth in its dependence on the directions \(\vec N\) --- the additional factors of \(x^i\) arise from converting the derivatives with respect to the ``rectangular'' coordinates \(x^i\) to derivatives with respect to the different directions (see \cref{eq:dd-derivative-eg} below for an example); in \cref{eq:dd-der-spatial}, these factors are provided instead by \(\Omh\) \cite{AH,Ash-in-Held,Geroch-asymp}.

The notion of regular direction-dependent tensors in any two coordinate charts in the same \(C^1\)-structure need not coincide \cite{AH, Herb-dd}. Thus, we are lead to restrict the differential structure of \(M\) at \(p\) to a so-called \(C^{>1}\)-structure, as follows.

\begin{definition}[\(C^{>1}\) differential structure]
\label{def:C>1-struc}
Consider any two \(C^1\) coordinate charts \(x^i\) and \(y^i\), in the same \(C^1\)-structure, containing the point \(p \in M\) such that for all \(i,j,k\) the \emph{transition functions}
\be\label{eq:C>1-struc}
    \frac{\partial^2 y^i(x)}{\partial x^j \partial x^k} \eqsp \frac{\partial^2 x^i(y)}{\partial y^j \partial y^k}
\ee
are regular direction-dependent at \(p\) in their respective coordinate charts. A collection of all coordinate charts related by \cref{eq:C>1-struc} defines a choice of \(C^{>1}\)-structure on \(M\) at \(p\).
\end{definition}

Given such a \(C^{>1}\)-structure at \(p\), any function whose derivatives upto the \((k-1)\)\textsuperscript{th} order vanish, whose \(k\)\textsuperscript{th} derivative is direction-independent, and whose \((k+1)\)\textsuperscript{th} derivative is regular direction-dependent will be called \(C^{>k}\). By a slight abuse of notation we denote regular direction-dependent functions by \(C^{>-1}\). Similarly, any tensor field is \(C^{>k}\) at \(p\) if all of its components in any coordinate chart in the chosen \(C^{>1}\)-structure are \(C^{>k}\) functions at \(p\).

As an example of direction-dependent tensors, consider Cartesian coordinates \((x,y)\) on \(\bb R^2\). Of course, \(\bb R^2\) can be given an analytic structure in terms of \((x,y)\). However, consider the polar coordinate functions \((r,\theta)\) defined in the standard way: \(x = r\cos\theta\) and \(y = r\sin\theta\), and let \(p : (x,y) = (0,0) \equiv r=0\) be the origin. The directions of approach to \(p\) are parametrised by \(\theta\). Then, 
\be
    \lim_{\to p} r = 0 \eqsp \lim_{\to p} d r = \cos\theta dx + \sin\theta dy
\ee 
and the function \(r\) is \(C^{>0}\) at \(p\). Similarly, \(\theta\) is \(C^{>-1}\) at \(p\), and we also have
\be
    \lim_{\to p} d \theta \text{ does not exist} \eqsp \lim_{\to p} r d\theta = -\sin\theta dx + \cos\theta dy 
\ee
and so the \(1\)-form \(r d\theta\) is \(C^{>-1}\) at \(p\). In general, any function \(f(x,y)\) such that \(\lim_{\to p} f(x,y)\) exists and depends only on \(\theta\) --- with \(\lim_{\to p} f(x,y) = \dd f(\theta)\) --- is direction-dependent. If \(f(x,y)\) is regular direction-dependent then the derivatives of \(\dd f(\theta)\) with respect to \(\theta\) to all orders exist and \cref{eq:dd-derivative} reads (with \(k=1\))
\be\label{eq:dd-derivative-eg}
    \lim_{\to p} x \frac{\partial f}{\partial x} = - \lim_{\to p} y \frac{\partial f}{\partial y} = - \sin\theta \cos\theta \frac{\partial \dd f}{\partial \theta}
\ee

\section{Rescaling function \(\Sigma\) in \(Ti^0\)}
\label{sec:Sigma}

In this section we show that the rescaling function \(\Sigma\) used in \cref{sec:null-dir} to construct the space \(\cyl\) of null and spatial directions exists for any spacetime satisfying \cref{def:AH}. Since the metric \(g_{ab}\) in the Ashtekar-Hansen structure is universal at \(i^0\), it induces a metric in the tangent space \(Ti^0\) which is isometric to the Minkowski metric. The constructions in the rest of this section only use the metric and not its connection and so we can safely carry them out in \(Ti^0\) as if we were on Minkowski spacetime.

We first briefly recall the Ashtekar-Hansen structure of Minkowski spacetime following \cite{AH}. In polar coordinates \((T,R,\theta^A)\) where \(\theta^A = (\theta,\phi)\) are the angular coordinates on \(\bb S^2\), the physical metric of Minkowski spacetime takes the form
\be
	\hat g_{ab} \equiv - dT^2 + dR^2 + R^2 s_{AB} d\theta^A d\theta^B
\ee
with the unit sphere metric \(s_{AB}\) given by
\be
    s_{AB} d\theta^A d\theta^B \equiv d\theta^2 + \sin^2\theta d\phi^2
\ee 
Defining new coordinates \((t,r)\) by
\be
    r+t = (R-T)^{-1} \eqsp r-t = (R+T)^{-1}
\ee
and the conformal factor \(\Omega = r^2 - t^2\) then gives the unphysical metric
\be\label{eq:Mink-g-unphys}\begin{split}
    g_{ab} = \Omega^2 \hat g_{ab} & \equiv - dt^2 + dr^2 + r^2 s_{AB} d\theta^A d\theta^B \\
    & = - dt^2 + dx^2 + dy^2 + dz^2
\end{split}\ee
where the Cartesian coordinates \((t,x,y,z)\) are defined in the usual way from \((t,r,\theta^A)\). Note that the Cartesian coordinates \((t,x,y,z)\) are \(C^1\), and the corresponding bases \((dt,dx,dy,dz)\) are \(C^0\), whereas the polar bases \((dr,rd\theta^A)\) are \(C^{>-1}\) at \(i^0\). The boundaries of this conformal-completion are then
\be
    \scri^\pm \equiv (t=\pm r, r > 0) \eqsp i^0 \equiv (t=0, r=0)
\ee
The vector field \(n^a\) is 
\be\label{eq:n-Ti0}\begin{split}
    n^a \defn \nabla^a \Omega & \equiv 2r \lb( \tfrac{t}{r}\partial_t + \partial_r \rb) \\
    & \equiv 2r \lb( \pm \partial_t + \partial_r \rb) \quad \text{on } t = \pm r \\
    & = 0 \quad \text{at } i^0
\end{split}\ee

To construct the hyperboloid \(\hyp\) of spatial directions, consider the coordinates \((\tau, \rho)\) defined by
\be\label{eq:hyp-coord}\begin{aligned}
	t & = \rho \sinh\tau \eqsp& r &= \rho \cosh\tau \\
    \rho & = (r^2 - t^2)^{\nfrac{1}{2}} \eqsp& \tau & = \tanh^{-1} \tfrac{t}{r} 
\end{aligned}\ee
so that
\be
    g_{ab} \equiv d\rho^2 + \rho^2 \lb( - d\tau^2 + \cosh^2\tau s_{AB} d\theta^A d\theta^B \rb) \eqsp \Omega = \rho^2
\ee
The \(C^{>-1}\) vector field
\be\label{eq:eta-Ti0}
    \dd\eta^a \defn \nabla^a \Omh \equiv \partial_\rho = \lb( 1 - \tfrac{t^2}{r^2} \rb)^{-\half} \lb( \tfrac{t}{r} \partial_t + \partial_r \rb) 
\ee
represents the unit spatial directions \(\vec\eta = (\tau ,\theta^A)\) in \(Ti^0\) with the space of \(\vec\eta\) being the hyperboloid \(\hyp\) given by \(\rho = 1\) as in \cref{fig:hyperboloid}. The induced metric on \(\hyp\) is
\be\label{eq:h-hyp-coord}
    \dd h_{ab} \equiv - d\tau^2 + \cosh^2\tau s_{AB} d\theta^A d\theta^B
\ee

Note that \(\dd\eta^a\) diverges as \(t \to \pm r\) and hence cannot represent the null directions at \(Ti^0\). To get both null and spatial directions we need a rescaling function as in \cref{def:Sigma} with the additional condition \cref{eq:Sigma-choice}. It suffices to show that at least one such rescaling function can be chosen in \(Ti^0\); the most general choice can be obtained using \cref{rem:freedom-Sigma,rem:freedom-Sigma-res}. It is easily seen that the choice
\be\label{eq:Sigma-r}
    \Sigma^{-1} = r
\ee
satisfies all the required conditions. On \(\hyp\) the function induced by \(\lim_{\to i^0}\Omh\Sigma\) is
\be\label{eq:Sigma-hyp}
    \dd \Sigma = \sech\tau = \lb( 1 - \tfrac{t^2}{r^2} \rb)^\half 
\ee
From \cref{eq:n-Ti0,eq:eta-Ti0} we have the rescaled null and spatial directions \(\vec N\) given by
\be\label{eq:N-Ti0}\begin{split}
    N^a = \tfrac{1}{2} \Sigma n^a & \equiv \lb( \tfrac{t}{r} \partial_t + \partial_r \rb) \\
    & \equiv \lb( \pm \partial_t + \partial_r \rb) \quad \text{in null directions} \\
    & = \dd\Sigma \dd\eta^a \quad \text{in spatial directions}
\end{split}\ee
The space of these rescaled null and spatial directions (with \cref{eq:Sigma-r}) is the cylinder \(\cyl\) given by \(r=1\) as in \cref{fig:cylinder}, and is parametrised by \(\vec N = (\alpha, \theta^A)\) where \(\alpha = t/r\). The boundaries \(\nulls^\pm \cong \bb S^2\) corresponding to \(\alpha = \pm 1\) represent the space of null directions.

The auxilliary normals \(L^a\) and \(l^a\) (\cref{eq:L-defn,eq:l-defn}) are then
\be
    L^a \equiv \tfrac{1}{2} \lb( \tfrac{t}{r} \partial_t - \partial_r \rb) \eqsp l^a \equiv \tfrac{1}{4} r^{-1} \lb( \tfrac{t}{r} \partial_t - \partial_r \rb)
\ee

To conformally-complete \(\hyp\), consider coordinate \(\beta \defn \sin^{-1} (\tanh\tau)\) with \(\beta \in (-\nfrac{\pi}{2}, \nfrac{\pi}{2})\) so that \(\dd\Sigma = \cos\beta\). Then, the manifold \(\tilde\hyp\) which includes the boundaries \(\beta = \pm \nfrac{\pi}{2}\) where \(\dd\Sigma = 0\) is a conformal-completion of \(\hyp\) with metric
\be\label{eq:h-beta}
	\tilde{\dd h}_{ab}  = \dd\Sigma^2 \dd h_{ab} \equiv - d\beta^2 +  s_{AB} d\theta^A d\theta^B
\ee
The diffeomorphism between the cylinder \(\cyl\) and the conformal-completion \(\tilde\hyp\) is given by the relation
\be
    (\alpha, \theta^A) = (\sin\beta, \theta^A) \text{ with } \nulls^\pm \equiv (\alpha = \pm 1, \theta^A) = (\beta = \pm \nfrac{\pi}{2}, \theta^A)
\ee
In the \((\alpha,\theta^A)\) coordinates on \(\cyl\) we have
\be\label{eq:h-alpha}\begin{split}
    \dd \Sigma & = (1-\alpha^2)^{\nfrac{1}{2}} \\
    \tilde{\dd h}_{ab} & \equiv - \frac{1}{1-\alpha^2} d\alpha^2 + s_{AB} d\theta^A d\theta^B
\end{split}\ee
Note that \(\tilde{\dd h}_{ab}\) does not extend smoothly  to \(\alpha = \pm 1\) and is \emph{not} the metric induced by \(g_{ab}\) on the \(\Sigma^{-1} = r = 1\) cylinder in \(Ti^0\). However, this still gives us the following useful fields on \(\cyl\).

The vector field \(\dd U^a\) (\cref{eq:U-defn}) can be computed to be
\be\label{eq:U-alpha}
    \dd U^a \equiv \sin\beta \partial_\beta \equiv \alpha (1-\alpha^2)^\half \partial_\alpha 
\ee
One can verify \cref{eq:U-defn,eq:U-on-N}, using \cref{eq:h-alpha}, with \(\dd\Sigma^{-1} \dd U^a \equiv \alpha \partial_\alpha \).

The pullback of the metric \(g_{ab}\) to \(\scri^\pm\) is
\be
    q_{ab} \equiv r^2 s_{AB} d\theta^A d\theta^B
\ee
The limit of the rescaled metric \(\tilde q_{ab} = \Sigma^2 q_{ab}\) induces the metric \(\tilde{\dd q}_{ab} \equiv s_{AB} d \theta^A d \theta^B\) on \(\nulls^\pm\). Similarly, on \(\cyl\) we have (from \cref{eq:h-alpha})
\be
    \lim_{\to \nulls^\pm} (\tilde{\dd h}_{ab} + \dd D_a \dd\Sigma \dd D_b \dd\Sigma) \equiv s_{AB} d \theta^A d \theta^B = \tilde{\dd q}_{ab}
\ee
It can be also be verified that the area element \cref{eq:nulls-area} induces the standard unit area element on \(\nulls^\pm\).

The reflection conformal isometry of \(\cyl\), \cref{eq:reflection-cyl}, is the map
\be
    \Upsilon : (\alpha, \theta^A) \mapsto (-\alpha, - \theta^A)
\ee
with \(\dd\varsigma = 1\) since our choice of \(\dd\Sigma\) (\cref{eq:h-alpha}) is invariant under this reflection. For arbitrary choices of \(\Sigma\) we would have, instead, \(\dd\varsigma = \dd\Sigma^{-1} (\Upsilon \circ \dd\Sigma) \).

\section{Relation to some coordinate-based approaches}
\label{sec:BS}

In this appendix we collect the relations between our covariant approach to some of the coordinate-based approaches used on Minkowski spacetime.

Given the Ashtekar-Hansen structure of \cref{def:AH}, consider a different choice of conformal factor \(\Omega_\bs = \varpi \Omega\) so that the Bondi condition \(\nabla_a^\bs \nabla_b^\bs \Omega_\bs\vert_{\scri^+} = 0\) holds. We denote quantities computed in this choice of conformal factor by a \(\bs\) for Bondi-Sachs. The Bondi-Sachs normal to \(\scri^+\) is then \(n^a_\bs \vert_{\scri^+} = \varpi^{-1} n^a\). The Bondi condition implies that \(\nabla_a^\bs n^a_\bs\vert_{\scri^+} = 0\) which in terms of the conformal-completion with \(\Omega\) gives \(\Lie_n \ln\varpi \vert_{\scri^+} = -\tfrac{1}{4} \nabla_a n^a \). From \cref{cond:Omega-at-i0} we have \(\nabla_a n^a\vert_{i^0} = 8\), and thus \(\varpi\vert_{\scri^+} = O(1/r)\), as we approach \(i^0\) along \(\scri^+\), where \(r\) is the \(C^{>0}\) ``radial'' coordinate at \(i^0\) from \cref{sec:Sigma}. The Bondi-Sachs parameter \(u\) on \(\scri^+\) is defined by \(n^a_\bs \nabla_a u = 1\) which gives \(u = O(1/r)\) with \(u \to - \infty\) being the limit to spatial infinity along \(\scri^+\). Note, however, that the unphysical metric in the Bondi-Sachs completion is \(g_{ab}^\bs = \varpi^2 g_{ab} = O( u^2) g_{ab}\) which diverges in any \(C^1\)-chart at \(i^0\) even on Minkowski spacetime. The asymptotic conditions at \(\scri^+\) in \cref{eq:F-asymp-null} correspond to the usual asymptotic expansions in the Bondi-Sachs coordinates (see, for example, \cite{CE}). 

From \cref{sec:Sigma}, any choice of the rescaling function behaves as \(\Sigma = O(1/r) = O(u)\) approaching \(i^0\) along \(\scri^+\). Converting to the conformal factor \(\Omega_\bs\), the null-regularity conditions of \cref{def:F-null-regular} imply that
\be\label{eq:F-null-regular-BS}
    \lim_{u \to -\infty} F_{ab} l^a_\bs n^b_\bs \eqsp \lim_{u \to - \infty} (*F)_{ab} l^a_\bs n^b_\bs
\ee
exist as smooth functions on \(\bb S^2\), while \cref{eq:J-E-null-regular} implies the falloffs
\be\label{eq:J-E-null-regular-BS}
    n_a^\bs J_\bs^a = O(1/u^{1+\epsilon}) \eqsp \mc E_a^\bs = O (1/u^{1+\epsilon})
\ee
for some small \(\epsilon > 0\). In the Newman-Penrose/Geroch-Held-Penrose notation \cite{NP,GHP}, \cref{eq:F-null-regular-BS,eq:J-E-null-regular-BS} imply that as \(u \to -\infty\), \(\varphi_2^\bs\) fallsoff as \(O(1/u^{1+\epsilon})\), \(\varphi_1^\bs\) has a limit as a smooth function on \(\bb S^2\), while we have not imposed any conditions on \(\varphi_0^\bs\). Thus our null-regularity conditions are consistent with the asymptotic expansions in the Bondi-Sachs parameter \(u\) on Minkowski spacetime used in \cite{CE}.

Similarly along spatial directions at \(i^0\), we can relate the hyperboloidal coordinate \(\rho\) (\cref{eq:hyp-coord}) to the Beig-Schmidt radial coordinate \cite{Beig-Schmidt} \(\rho_\bs\) by \(\rho_\bs = O(\rho^{-1})\) (where now we use \(\bs\) for Beig-Schmidt). The requirement that the unphysical metric \(g_{ab}\) is \(C^{>0}\) at \(i^0\) translates to, in terms of the physical metric \(\hat g_{ab}\)
\be
    \hat g_{ab} = \hat \eta_{ab} + \frac{\hat g^{(1)}_{ab}(\tau,\theta^A)}{\rho_\bs} + o(1/\rho_\bs)
\ee
where the tensor components are evaluated in an asymptotic Cartesian chart for the asymptotic Minkowski metric \(\hat\eta_{ab}\), and \(\hat g^{(1)}_{ab}\) are smooth functions on the hyperboloid \(\hyp\). Similarly, it can be verified that the asymptotic conditions in \cref{eq:F-asymp-spatial} in this asymptotic chart become
\be
    F_{ab} = \frac{ F_{ab}^{(2)}(\tau,\theta^A)}{\rho_\bs^2} + o(1/\rho_\bs^2)  \eqsp \hat J^a = o(1/\rho_\bs^3)
\ee
which are consistent with the leading order asymptotic expansions used in \cite{Beig-Schmidt, CE}. On Minkowski spacetime, using the explicit transformations between the Bondi-Sachs and Beig-Schmidt coordinates reproduces the analysis of \cite{CE}. But, as mentioned in the Introduction, the explicit transformations from the Bondi-Sachs to Beig-Schmidt coordinates are not available in more general spacetimes, and our covariant approach is more suited to the general matching problem.
 
It would be of interest to relate our formalism to the one used by Friedrich \cite{Friedrich} to treat spatial infinity and the related coordinates used in \cite{HT-em,Tro}.

\section{Solutions to Maxwell equation on \(\hyp\) and their extensions to \(\cyl\)}
\label{sec:wave-hyp}

In this section we examine solutions to the Maxwell equation \cref{eq:Max-hyp} on the unit-hyperboloid \(\hyp\), see also \cite{CE,HT-em}. We will focus on the electric field \(\dd E_a\), the analysis for the magnetic field \(\dd B_a\) is completely analogous. 

The Maxwell equations on \(\hyp\) \cref{eq:Max-hyp} imply that there exists a potential \(\dd V\) for \(\dd E_a\) so that
\be\label{eq:wave-hyp}
    \dd E_a = \dd D_a \dd V \eqsp \dd D^a \dd D_a \dd V = 0
\ee
In the coordinates \((\alpha = \tanh\tau, \theta^A)\) the metric \cref{eq:h-hyp-coord} on \(\hyp\) is
\be
    \dd h_{ab} \equiv - \frac{1}{(1-\alpha^2)^2} d\alpha^2 + \frac{1}{1-\alpha^2} s_{AB} d\theta^A d\theta^B
\ee
and we can solve \cref{eq:wave-hyp} using a decomposition \(\dd V = \sum\limits_{\ell,m} V_{\ell,m}(\alpha)Y_{\ell, m}(\theta^A)\) where \(Y_{\ell, m}(\theta^A)\) are the spherical harmonic functions and each \(V_{\ell,m}(\alpha)\) satisfies
\be\label{eq:red-wave-hyp}
    (1-\alpha^2)\frac{d^2}{d\alpha^2} V_{\ell,m} + \ell(\ell+1) V_{\ell,m} = 0
\ee
This admits solutions \(V_{\ell,m}\) as linear combinations of\footnote{Alternatively, \cref{eq:red-wave-hyp} can also be solved in terms of the \emph{Gau\ss\ hypergeometric functions} \({}_2F_1(- \tfrac{\ell+1}{2},\tfrac{\ell}{2};\tfrac{1}{2};\alpha^2)\) and \(\alpha \times {}_2F_1(- \tfrac{\ell}{2},\tfrac{\ell+1}{2};\tfrac{3}{2};\alpha^2)\) which do not miss the linear in \(\alpha\) solution but somewhat obscure the parity transformation under \(\Upsilon\). These can be related to the solutions \cref{eq:Legendre-soln} using the transformation formulae in \S~3.2 \cite{special-func} or \S~14.3 \cite{DLMF}.}
\be\label{eq:Legendre-soln}
    (1-\alpha^2)^\half P^1_\ell (\alpha) \eqsp (1-\alpha^2)^\half Q^1_\ell(\alpha)
\ee
where \(P^1_\ell\) and \(Q^1_\ell\) are the \emph{Legendre functions} \cite{special-func}. Note, that for \(\ell = 0\) these miss out the obvious solution \(V_{\ell=0,m=0} \propto \alpha\). Under the time reflection isometry \(\alpha \mapsto -\alpha\) on \(\hyp\), the solutions spanned by \(P^1_\ell\) and \(Q^1_\ell\) have a parity of \((-1)^{\ell+1}\) and \((-1)^\ell\), respectively. Combined with the well-known parity \((-1)^\ell\) of the spherical harmonics \(Y_{\ell,m}\) under \(\theta^A \mapsto -\theta^A\), we get two linearly independent solutions \(\dd V^{\rsub{(odd)}}\) and \(\dd V^{\rsub{(even)}}\) to \cref{eq:wave-hyp} which are odd and even, respectively, under the reflection isometry \(\Upsilon: (\alpha, \theta^A) \mapsto (-\alpha, -\theta^A)\) of \(\hyp\).

These solutions have the following behaviour in \(\alpha\) (see \S~3.9.2 \cite{special-func} or \S~14.8 \cite{DLMF})
\be\label{eq:wave-soln}\begin{split}
    \ell = 0 : &\quad
    \begin{aligned}
        \dd V^{\rsub{(odd)}} & = \text{constant} \times \alpha \\
        \dd V^{\rsub{(even)}} & = \text{constant} 
    \end{aligned} \\[1.5ex]
    \ell \neq 0 : &\quad
    \begin{aligned}
        \dd V^{\rsub{(odd)}} & = V_1(\theta^A)(1-\alpha^2) + O\lb((1-\alpha^2)^2\rb) \\
        \dd V^{\rsub{(even)}} & = V_2(\theta^A) + O\lb((1-\alpha^2)\ln (1-\alpha^2)\rb) 
    \end{aligned}
\end{split}\ee
From the electric field \(\dd E_a = \dd D_a \dd V\) we see that, for \(\ell = 0\), \(\dd V^{\rsub{(even)}}\) is a pure-gauge solution (with vanishing electric field), while \(\dd V^{\rsub{(odd)}}\) is the Coulomb solution (with a static electric field). The Coulomb solution seems to have been missed in \cite{CE,HT-em} but this does not affect their results; for \(\ell \neq 0\) the above asymptotics in \(\alpha\) reproduce those obtained by \cite{CE, HT-em}.

Using the diffeomorphism between \(\hyp\) and \(\cyl \setminus \nulls^\pm\) we can treat \(\dd V\) and \(\dd E_a\) as fields on \(\cyl\). For null-regular Maxwell fields we need \(\dd\Sigma^{-1} \dd E_a \dd U^a\) to have a finite limit to \(\nulls^\pm\), where \(\alpha = \pm 1\). From \cref{eq:h-alpha,eq:U-alpha,eq:wave-soln} we see that \(\dd\Sigma^{-1} \dd E_a \dd U^a = \alpha \partial_\alpha \dd V\) has a finite limit to \(\nulls^\pm\) only for the solutions \(\dd V = \dd V^{\rsub{(odd)}}\) (and trivially for the pure-gauge solution). Thus, under the reflection \(\Upsilon\), for null-regular Maxwell fields, we have 
\be\begin{split}
    \Upsilon \circ (\dd\Sigma^{-1} \dd E_a \dd U^a)\vert_{\nulls^-} & = - (\dd\Sigma^{-1} \dd E_a \dd U^a)\vert_{\nulls^+}
\end{split}\ee 
in the choice rescaling function \(\Sigma^{-1} = r\) from \cref{sec:Sigma}; for more general choices we get \cref{eq:reflection-electric}.



\bibliographystyle{JHEP}
\bibliography{BMS-matching}      

\providecommand{\href}[2]{#2}\begingroup\raggedright\begin{thebibliography}{10}

\bibitem{BBM}
H.~Bondi, M.~G.~J. van~der Burg and A.~W.~K. Metzner, \emph{{Gravitational
  waves in general relativity. 7. Waves from axisymmetric isolated systems}},
  \href{https://doi.org/10.1098/rspa.1962.0161}{\emph{Proc. Roy. Soc. Lond.}
  {\bfseries A269} (1962) 21}.

\bibitem{Sachs1}
R.~K. Sachs, \emph{{Gravitational waves in general relativity. 8. Waves in
  asymptotically flat space-times}},
  \href{https://doi.org/10.1098/rspa.1962.0206}{\emph{Proc. Roy. Soc. Lond.}
  {\bfseries A270} (1962) 103}.

\bibitem{Sachs2}
R.~Sachs, \emph{{Asymptotic symmetries in gravitational theory}},
  \href{https://doi.org/10.1103/PhysRev.128.2851}{\emph{Phys. Rev.} {\bfseries
  128} (1962) 2851}.

\bibitem{Penrose}
R.~Penrose, \emph{Zero rest-mass fields including gravitation: asymptotic
  behaviour}, \href{https://doi.org/10.1098/rspa.1965.0058}{\emph{Proc. R. Soc.
  A} {\bfseries 284} (1965) 159}.

\bibitem{GW}
R.~Geroch and J.~Winicour, \emph{Linkages in general relativity},
  \href{https://doi.org/10.1063/1.524987}{\emph{J. Math. Phys.} {\bfseries 22}
  (1981) 803}.

\bibitem{AS-symp}
A.~Ashtekar and M.~Streubel, \emph{{Symplectic Geometry of Radiative Modes and
  Conserved Quantities at Null Infinity}},
  \href{https://doi.org/10.1098/rspa.1981.0109}{\emph{Proc. Roy. Soc. Lond.}
  {\bfseries A376} (1981) 585}.

\bibitem{WZ}
R.~M. Wald and A.~Zoupas, \emph{{A General definition of `conserved quantities'
  in general relativity and other theories of gravity}},
  \href{https://doi.org/10.1103/PhysRevD.61.084027}{\emph{Phys. Rev.}
  {\bfseries D61} (2000) 084027}
  [\href{https://arxiv.org/abs/gr-qc/9911095}{{\ttfamily gr-qc/9911095}}].

\bibitem{ADM}
R.~Arnowitt, S.~Deser and C.~W. Misner, \emph{{The Dynamics of General
  Relativity}},  in \emph{{Gravitation: An Introduction to Current Research}}
  (L.~Witten, ed.).
\newblock Wiley, New York, 1962.

\bibitem{ADMG}
R.~Geroch, \emph{{Structure of the Gravitational Field at Spatial Infinity}},
  \href{https://doi.org/10.1063/1.1666094}{\emph{J. Math. Phys.} {\bfseries 13}
  (1972) 956}.

\bibitem{CR}
A.~Corichi and J.~D. Reyes, \emph{{The gravitational Hamiltonian, first order
  action, Poincaré charges and surface terms}},
  \href{https://doi.org/10.1088/0264-9381/32/19/195024}{\emph{Class. Quant.
  Grav.} {\bfseries 32} (2015) 195024}
  [\href{https://arxiv.org/abs/1505.01518}{{\ttfamily 1505.01518}}].

\bibitem{Beig-Schmidt}
R.~Beig and B.~G. Schmidt, \emph{Einstein's equations near spatial infinity},
  \href{https://doi.org/10.1007/BF01211056}{\emph{Comm. Math. Phys.} {\bfseries
  87} (1982) 65}.

\bibitem{AH}
A.~Ashtekar and R.~O. Hansen, \emph{{A unified treatment of null and spatial
  infinity in general relativity. I. Universal structure, asymptotic
  symmetries, and conserved quantities at spatial infinity}},
  \href{https://doi.org/10.1063/1.523863}{\emph{J. Math. Phys.} {\bfseries 19}
  (1978) 1542}.

\bibitem{Sommers}
P.~Sommers, \emph{The geometry of the gravitational field at spacelike
  infinity}, \href{https://doi.org/10.1063/1.523698}{\emph{J. Math. Phys.}
  {\bfseries 19} (1978) 549}.

\bibitem{Ash-in-Held}
A.~Ashtekar, \emph{{Asymptotic Structure of the Gravitational Field at Spatial
  Infinity}},  in \emph{{General Relativity and Gravitation. One Hundered Years
  After the Birth of Albert Einstein}} (A.~Held, ed.), vol.~2, pp.~37--69.
\newblock Plenum Press, New York, 1980.

\bibitem{Ash-Rom}
A.~Ashtekar and J.~D. Romano, \emph{Spatial infinity as a boundary of
  spacetime}, \href{https://doi.org/10.1088/0264-9381/9/4/019}{\emph{Class.
  Quant. Grav.} {\bfseries 9} (1992) 1069}.

\bibitem{Friedrich}
H.~Friedrich, \emph{Gravitational fields near space-like and null infinity},
  \href{https://doi.org/10.1016/S0393-0440(97)82168-7}{\emph{J. Geom. Phys.}
  {\bfseries 24} (1998) 83}.

\bibitem{AES}
A.~Ashtekar, J.~Engle and D.~Sloan, \emph{{Asymptotics and Hamiltonians in a
  First order formalism}},
  \href{https://doi.org/10.1088/0264-9381/25/9/095020}{\emph{Class. Quant.
  Grav.} {\bfseries 25} (2008) 095020}
  [\href{https://arxiv.org/abs/0802.2527}{{\ttfamily 0802.2527}}].

\bibitem{Geroch-asymp}
R.~Geroch, \emph{{Asymptotic structure of space-time}},  in \emph{{Asymptotic
  structure of space-time}} (F.~P. Esposito and L.~Witten, eds.).
\newblock Plenum Press, New York, 1977.

\bibitem{Ash-Mag-Ash}
A.~Ashtekar and A.~Magnon-Ashtekar, \emph{{Energy-Momentum in General
  Relativity}}, \href{https://doi.org/10.1103/PhysRevLett.43.181}{\emph{Phys.
  Rev. Lett.} {\bfseries 43} (1979) 181} errata:
  \href{http://dx.doi.org/10.1103/PhysRevLett.43.649}{{\it Phys. Rev. Lett.}
  {\bf 43} (1979) 649}.

\bibitem{AS-ang-mom}
A.~Ashtekar and M.~Streubel, \emph{On angular momentum of stationary
  gravitating systems}, \href{https://doi.org/10.1063/1.524242}{\emph{J. Math.
  Phys.} {\bfseries 20} (1979) 1362}.

\bibitem{HL-GR-matching}
M.~Herberthson and M.~Ludvigsen, \emph{A relationship between future and past
  null infinity}, \href{https://doi.org/10.1007/BF00756992}{\emph{Gen. Rel.
  Grav.} {\bfseries 24} (1992) 1185}.

\bibitem{Stro-CK-match}
A.~Strominger, \emph{{On BMS Invariance of Gravitational Scattering}},
  \href{https://doi.org/10.1007/JHEP07(2014)152}{\emph{JHEP} {\bfseries 07}
  (2014) 152} [\href{https://arxiv.org/abs/1312.2229}{{\ttfamily 1312.2229}}].

\bibitem{HLMS}
T.~He, V.~Lysov, P.~Mitra and A.~Strominger, \emph{{BMS supertranslations and
  Weinberg’s soft graviton theorem}},
  \href{https://doi.org/10.1007/JHEP05(2015)151}{\emph{JHEP} {\bfseries 05}
  (2015) 151} [\href{https://arxiv.org/abs/1401.7026}{{\ttfamily 1401.7026}}].

\bibitem{SZ}
A.~Strominger and A.~Zhiboedov, \emph{{Gravitational Memory, BMS
  Supertranslations and Soft Theorems}},
  \href{https://doi.org/10.1007/JHEP01(2016)086}{\emph{JHEP} {\bfseries 01}
  (2016) 086} [\href{https://arxiv.org/abs/1411.5745}{{\ttfamily 1411.5745}}].

\bibitem{HPS}
S.~W. Hawking, M.~J. Perry and A.~Strominger, \emph{{Soft Hair on Black
  Holes}}, \href{https://doi.org/10.1103/PhysRevLett.116.231301}{\emph{Phys.
  Rev. Lett.} {\bfseries 116} (2016) 231301}
  [\href{https://arxiv.org/abs/1601.00921}{{\ttfamily 1601.00921}}].

\bibitem{Hawking}
S.~W. Hawking, \emph{{The Information Paradox for Black Holes}},  2015,
  \href{https://arxiv.org/abs/1509.01147}{{\ttfamily 1509.01147}}.

\bibitem{BP}
R.~Bousso and M.~Porrati, \emph{{Soft Hair as a Soft Wig}},
  \href{https://doi.org/10.1088/1361-6382/aa8be2}{\emph{Class. Quant. Grav.}
  {\bfseries 34} (2017) 204001}
  [\href{https://arxiv.org/abs/1706.00436}{{\ttfamily 1706.00436}}].

\bibitem{CK}
D.~Christodoulou and S.~Klainerman, \emph{{The global nonlinear stability of
  the Minkowski space}}. Princeton University Press, 1993.

\bibitem{Ash-CK-triv}
A.~Ashtekar, \emph{{The BMS group, conservation laws, and soft gravitons}},  8
  Nov. 2016.
\newblock Talk presented at the Perimeter Institute for Theoretical Physics.
  Available online at \url{http://pirsa.org/16080055/}.

\bibitem{CE}
M.~Campiglia and R.~Eyheralde, \emph{{Asymptotic $U(1)$ charges at spatial
  infinity}}, \href{https://doi.org/10.1007/JHEP11(2017)168}{\emph{JHEP}
  {\bfseries 11} (2017) 168}
  [\href{https://arxiv.org/abs/1703.07884}{{\ttfamily 1703.07884}}].

\bibitem{HT-em}
M.~Henneaux and C.~Troessaert, \emph{{Asymptotic symmetries of electromagnetism
  at spatial infinity}},  \href{https://arxiv.org/abs/1803.10194}{{\ttfamily
  1803.10194}}.

\bibitem{AES-p-form}
H.~Afshar, E.~Esmaeili and M.~M. Sheikh-Jabbari, \emph{{Asymptotic Symmetries
  in $p$-Form Theories}},
  \href{https://doi.org/10.1007/JHEP05(2018)042}{\emph{JHEP} {\bfseries 05}
  (2018) 042} [\href{https://arxiv.org/abs/1801.07752}{{\ttfamily
  1801.07752}}].

\bibitem{Tro}
C.~Troessaert, \emph{{The BMS4 algebra at spatial infinity}},
  \href{https://doi.org/10.1088/1361-6382/aaae22}{\emph{Class. Quant. Grav.}
  {\bfseries 35} (2018) 074003}
  [\href{https://arxiv.org/abs/1704.06223}{{\ttfamily 1704.06223}}].

\bibitem{Hawking-Ellis}
S.~Hawking and G.~Ellis, \emph{{The Large scale structure of space-time}}.
  Cambridge University Press, London-New York, 1973.

\bibitem{Herb-Kerr}
M.~Herberthson, \emph{{A \(C^{>1}\) Completion of the Kerr Space-Time at
  Spacelike Infinity Including \(I^+\) and \(I^-\)}},
  \href{https://doi.org/10.1023/A:1012085301675}{\emph{Gen. Rel. Grav.}
  {\bfseries 33} (2001) 1197}.

\bibitem{Berg}
P.~G. Bergmann, \emph{{``Gauge-Invariant'' Variables in General Relativity}},
  \href{https://doi.org/10.1103/PhysRev.124.274}{\emph{Phys. Rev.} {\bfseries
  124} (1961) 274}.

\bibitem{Ash-log}
A.~Ashtekar, \emph{Logarithmic ambiguities in the description of spatial
  infinity}, \href{https://doi.org/10.1007/BF01889278}{\emph{Found. Phys.}
  {\bfseries 15} (1985) 419}.

\bibitem{Chr-log}
P.~T. Chruściel, \emph{{On the structure of spatial infinity. II. Geodesically
  regular Ashtekar–Hansen structures}},
  \href{https://doi.org/10.1063/1.528209}{\emph{J. Math. Phys.} {\bfseries 30}
  (1989) 2094}.

\bibitem{Wald-book}
R.~M. Wald, \emph{{General Relativity}}. The University of Chicago Press, 1984.

\bibitem{Harris}
J.~Harris, \emph{Algebraic Geometry: A First Course}, vol.~133 of
  \emph{Graduate Texts in Mathematics}. Springer-Verlag, New York, 1~ed., 1992.

\bibitem{Herb-dd}
M.~Herberthson, \emph{{On the differentiability conditions at space-like
  infinity}}, \href{https://doi.org/10.1088/0264-9381/15/12/016}{\emph{Class.
  Quant. Grav.} {\bfseries 15} (1998) 3873}
  [\href{https://arxiv.org/abs/gr-qc/9712058}{{\ttfamily gr-qc/9712058}}].

\bibitem{Porrill}
J.~Porrill, \emph{The structure of timelike infinity for isolated systems},
  \href{https://doi.org/10.1098/rspa.1982.0075}{\emph{Proc. R. Soc. A}
  {\bfseries 381} (1982) 323}.

\bibitem{Cutler}
C.~Cutler, \emph{Properties of spacetimes that are asymptotically flat at
  timelike infinity},
  \href{https://doi.org/10.1088/0264-9381/6/8/009}{\emph{Class. Quant. Grav.}
  {\bfseries 6} (1989) 1075}.

\bibitem{HL-timelike}
M.~Herberthson and M.~Ludvigsen, \emph{Time-like infinity and
  direction-dependent metrics},
  \href{https://doi.org/10.1088/0264-9381/11/1/019}{\emph{Class. Quant. Grav.}
  {\bfseries 11} (1994) 187}.

\bibitem{CFP}
V.~Chandrasekaran, {\'E}.~{\'E}. Flanagan and K.~Prabhu, \emph{Symmetries and
  charges of general relativity at null boundaries},
  \href{https://arxiv.org/abs/1807.11499}{{\ttfamily 1807.11499}}.

\bibitem{Stro-YM}
A.~Strominger, \emph{{Asymptotic Symmetries of Yang-Mills Theory}},
  \href{https://doi.org/10.1007/JHEP07(2014)151}{\emph{JHEP} {\bfseries 07}
  (2014) 151} [\href{https://arxiv.org/abs/1308.0589}{{\ttfamily 1308.0589}}].

\bibitem{NP-red}
E.~T. Newman and R.~Penrose, \emph{{Note on the Bondi-Metzner-Sachs Group}},
  \href{https://doi.org/10.1063/1.1931221}{\emph{J. Math. Phys.} {\bfseries 7}
  (1966) 863}.

\bibitem{RT-parity}
T.~Regge and C.~Teitelboim, \emph{{Role of Surface Integrals in the Hamiltonian
  Formulation of General Relativity}},
  \href{https://doi.org/10.1016/0003-4916(74)90404-7}{\emph{Annals Phys.}
  {\bfseries 88} (1974) 286}.

\bibitem{HT}
M.~Henneaux and C.~Troessaert, \emph{{BMS Group at Spatial Infinity: the
  Hamiltonian (ADM) approach}},
  \href{https://doi.org/10.1007/JHEP03(2018)147}{\emph{JHEP} {\bfseries 03}
  (2018) 147} [\href{https://arxiv.org/abs/1801.03718}{{\ttfamily
  1801.03718}}].

\bibitem{NP}
E.~Newman and R.~Penrose, \emph{{An Approach to Gravitational Radiation by a
  Method of Spin Coefficients}},
  \href{https://doi.org/10.1063/1.1724257}{\emph{J. Math. Phys.} {\bfseries 3}
  (1962) 566} errata: \href{http://dx.doi.org/10.1063/1.1704025}{{\it J. Math.
  Phys.} {\bf 4} (1963) 998}.

\bibitem{GHP}
R.~P. Geroch, A.~Held and R.~Penrose, \emph{{A space-time calculus based on
  pairs of null directions}}, \href{https://doi.org/10.1063/1.1666410}{\emph{J.
  Math. Phys.} {\bfseries 14} (1973) 874}.

\bibitem{special-func}
A.~Erd{\'e}lyi, W.~Magnus, F.~Oberhettinger and F.~G. Tricomi, \emph{Higher
  Transcendental Functions. {V}ol. {I}}. McGraw-Hill Book Company, Inc., New
  York-Toronto-London, 1953.

\bibitem{DLMF}
``{\it NIST Digital Library of Mathematical Functions}.''
  \url{http://dlmf.nist.gov/}, Release 1.0.18 of 2018-03-27.
\newblock F.~W.~J. Olver, A.~B. {Olde Daalhuis}, D.~W. Lozier, B.~I. Schneider,
  R.~F. Boisvert, C.~W. Clark, B.~R. Miller and B.~V. Saunders, eds.

\end{thebibliography}\endgroup
\end{document}